\documentclass[]{emulateapj}

\usepackage{natbib}
\newcommand{\eqref}[1]{(\ref{#1})}

\shorttitle{}
\shortauthors{}

\begin{document}

\title{The Influence of Concentration and Dynamical State on Scatter in the Galaxy Cluster Mass-Temperature Relation}

\author{Hsiang-Yi Karen Yang\altaffilmark{1},
Paul M. Ricker\altaffilmark{1,2},
and P.\ M. Sutter\altaffilmark{3}}
\altaffiltext{1}{Department of Astronomy, University of Illinois, Urbana, IL}
\altaffiltext{2}{National Center for Supercomputing Applications, Urbana, IL}
\altaffiltext{3}{Department of Physics, University of Illinois, Urbana, IL}
\email{hyang20@illinois.edu, pmricker@illinois.edu, psutter2@illinois.edu}

\begin{abstract}

Using a hydrodynamics plus $N$-body simulation of galaxy cluster formation
within a large volume and mock Chandra X-ray observations, we study the form and evolution of the intrinsic scatter about the best-fit X-ray temperature-mass relation for clusters. We investigate the physical origin of the scatter by correlating it with quantities that are closely related to clusters' formation and merging histories. We also examine the distribution of the scatter for merging and nonmerging populations, identified  using halo merger trees derived from the simulation as well as X-ray substructure measures. We find a strong correlation between the scatter in the $M-T_X$ relation and the halo concentration, in the sense that more concentrated clusters tend to be cooler than clusters with similar masses. No bias is found between the merging and relaxed clusters, but merging clusters generally have greater scatter, which is related to the properties of the distribution of halo concentrations. We also detect a signature of non-lognormality in the distribution of scatter for our simulated clusters both at $z=0$ and at $z=1$. A detailed comparison of merging clusters identified by substructure measures and by halo merger trees is given in the discussion. We conclude that, when cooling-related effects are neglected, the variation in halo concentrations is a more important factor for driving the intrinsic scatter in the $M-T_X$ relation, while departures from hydrostatic equilibrium due to cluster mergers have a minor effect.

\end{abstract}



\section{Introduction}

Galaxy clusters are potentially valuable cosmological probes because of their unique 
position in the hierarchy of structure formation.
They are the largest gravitationally bound objects,
having just separated from the cosmic expansion and collapsed from density 
fluctuations in the
past few billion years. Therefore, statistical measures of clusters,
such as their mass distribution
as a function of redshift, are sensitive to the cosmic matter density parameter 
$\Omega_m$, the dark energy density parameter $\Omega_{de}$,
the normalization of the primordial fluctuation spectrum $\sigma_8$, and the dark
energy equation of state parameter $w$ \citep{2005astro.ph..7013H}.
Future cluster surveys will yield cosmological parameter constraints that are complementary to those from upcoming microwave
background probes (e.g.\ Planck) and Type~Ia supernova observations.

However, clusters present us with a dilemma: their masses are well-predicted by numerical
simulations, but in the real world, 80--85\% of their mass is in the form of invisible dark matter.
In order to make contact with observations, observational mass proxies are needed. Fortunately,
cluster masses correlate with many observable quantities, such as X-ray temperature $T_X$,
X-ray luminosity $L_X$, optical richness, infrared luminosity, and the Sunyaev-Zel'dovich effect
\citep{1999ApJ...517..627M,2003ApJ...591..749L,2005A&A...433..431P,2005ApJ...633..122H,2006ApJ...648..956S}.
These mass-observable relations indicate that clusters are fairly regular and close to equilibrium,
despite having dynamical timescales of order 1/10 the age of the universe.

For clusters to provide meaningful constraints on the cosmological parameters,
the systematic errors
in mass estimates based on these relations must be well-understood. For example, masses
determined under the assumption of hydrostatic equilibrium in general underestimate the spherical
overdensity mass $M_{200}$ by $\sim20\%$ \citep{2007ApJ...655...98N}.
The X-ray temperature determined through spectral fitting
is also known to bias low with respect to the emission-weighted
temperature commonly used in numerical simulations \citep{2005ApJ...618L...1R}.
The discrepancy in the normalization of the $M$--$T_X$ relation between observations and numerical
simulations has been alleviated only recently by taking these two effects into account. This
emphasizes
the importance of using mock-observation tools to make direct comparisons between results from
numerical simulations and observational data.

The origin and distribution of scatter in these relations are also important. Due to the exponential
shape of the mass function, scatter in the $M$--$X$ relation (for observable $X$)
boosts the number density
of clusters observed in logarithmic bins of $X$, as the overall number of lower-mass
clusters scattering to higher values of $X$ far exceeds the number of high-mass clusters scattering
in the opposite direction. Underestimating this scatter can lead to an overestimate of $\sigma_8$, for
instance \citep{2002ApJ...577..579R}.
Attempts have been made to reduce the observed scatter to get better constraints on
cluster masses. Such attempts include the use of core-excised quantities to reduce the large scatter
in the $L_X$--$T_X$ relation due to the effects of cool cores
\citep{1998MNRAS.297L..57A,2006ApJ...639...64O} and the invention of the X-ray
counterpart of the Compton y-parameter, $Y_X \equiv M_{\rm gas}T_X$, to obtain a very tight $M$--$Y_X$
correlation \citep{2006ApJ...650..128K}.
These successful examples illustrate the possibility of obtaining better mass
estimates if our knowledge of the physical origin of scatter is improved. It is also possible
to self-calibrate cluster surveys
\citep{2002ApJ...577..569L,2004ApJ...613...41M,2004PhRvD..70d3504L},
fitting the mass-observable relation as an unknown together with the cosmological parameters.
However, this technique requires assumptions about the functional form and the mass and redshift
dependence of the scatter, and errors in these assumptions can lead to
misinterpretation of the obtained constraints \citep{2005PhRvD..72d3006L}.

In this paper we begin a systematic study of the influence of internal cluster physics
on the scatter in cluster mass-observable relations. We focus initially on X-ray
observables, as they are less directly dependent on the uncertain details of galaxy
formation than, for example, optical observables. Our aim in this paper is to
examine the effects of mergers and dynamical state in isolation from effects due
to radiative cooling, feedback due to stars and black holes, and diffusive transport.
Future papers will examine these other effects separately.
Accordingly, the simulation described here includes only dark matter
and gasdynamics.
While we examine substructure-based measures of dynamical state to
make contact with previous work, we extend this work by considering
direct measurements of dynamical state that may not be observable but that
can give us physical insight into the origin and form of the scatter. For example, we use
halo merger histories derived from halo catalogs created every $100 h^{-1}$\ Myr to
identify which clusters are merging at any given epoch. \S~\ref{Sec:simulations}
describes our numerical methods and simulation parameters. In \S~\ref{Sec:analysis}
we describe our merger tree and virial analysis procedures, our method for
generating simulated X-ray observations, and the substructure measures we employ.
We present our results in \S~\ref{Sec:results} and discuss them in \S~\ref{Sec:discussion}.
Finally, we summarize our conclusions in \S~\ref{Sec:conclusion}.

Throughout this paper we have taken the Hubble constant to be $H_0 = 100h$~km~s$^{-1}$~Mpc$^{-1}$, with $h = 0.708$.  When quoting masses or radii defined using an overdensity criterion
(e.g., $M_{200}$, $R_{200}$), we refer to overdensities relative to the critical density
at the relevant epoch.


\section{Cosmological simulation}
\label{Sec:simulations}

\subsection{Numerical methods}

The simulation described here was performed
using FLASH, an Eulerian hydrodynamics plus $N$-body code
originally developed for simulations of Type~Ia supernovae and related phenomena
\citep{FLASH}. The flexibility of FLASH's application framework has enabled it to be
applied to a wide range of problems.  In the process it has been extensively validated,
both for hydrodynamical \citep{2002ApJS..143..201C} and cosmological $N$-body \citep{2005ApJS..160...28H, 2008CS&D....1a5003H} applications. We used version 2.4 of FLASH
together with the local transform-based multigrid Poisson solver described by \cite{2008ApJS..176..293R}. The Euler equations describing the behavior of the intracluster
medium (ICM) were
solved using the Piecewise-Parabolic Method (PPM); extensive details of the FLASH
implementation of PPM are given by \cite{FLASH}.  The $N$-body component describing the
behavior of the dark matter was handled using the particle-mesh technique with cloud-in-cell
interpolation.  Because we are concerned in this paper only with the effect of gravity-driven
variations in dynamical state on mass-observable relations, the calculation described here
did not employ radiative cooling or feedback due to star formation or active galaxies.

\subsection{Simulation details}

The results presented here are based on a FLASH simulation of structure formation in
the $\Lambda$CDM cosmology within a 3D cubical volume spanning $256 h^{-1}$~Mpc.
Initial conditions were generated for a starting redshift $z$ of 66 using GRAFIC
\citep{grafic} with an initial power spectrum generated using CMBFAST \citep{cmbfast}.
The cosmological parameter values used were chosen to be consistent with the third-year
WMAP results \citep{wmap3}: present-day matter density parameter $\Omega_{m0} = 0.262$,
present-day baryonic density parameter $\Omega_{b0} = 0.0437$, present-day cosmological
constant density parameter $\Omega_{\Lambda0} = 0.738$, and matter power spectrum
normalization $\sigma_8 = 0.74$. The simulation contains $1024^3$ dark matter particles
with a particle mass $m_p = 9.2\times10^8 h^{-1} M_\odot$. The mesh used for the gasdynamics
and potential solution was fully refined to $1024^3$ zones, which corresponds to a
zone spacing of $250 h^{-1}$~kpc. Considering the effect of resolution on the computed
abundances of halos of different mass \citep{2007ApJ...671.1160L},
with these parameters we are able to capture all
halos containing more than 3150 particles (i.e. total mass $2.9\times10^{12} h^{-1} M_\odot$)
and 1150 particles (i.e. $1.1\times10^{12} h^{-1} M_\odot$) at $z = 0$ and $z = 1$,
respectively. The halos are identified using the friends-of-friends (FOF) algorithm. The overdensity
mass and radius, $M_\Delta$ and $R_\Delta$, are then found by growing spheres around each FOF center until the averaged total density is $\Delta$ times the critical density of the universe.

The simulation was carried out using 800 processors of the Cray XT4 system at Oak Ridge
National Laboratory, requiring a total of 16,500 CPU-hours. Gas and particle snapshots were
written to disk every
$100 h^{-1}$~Myr beginning at $z = 2$, yielding a total of 117 snapshots containing
15~TB of data.


\section{Analysis of the simulation}
\label{Sec:analysis}

\subsection{Merger tree analysis}

In order to directly quantify the dynamical state of clusters without relying on morphology, we generate merger trees for each cluster in our simulation and find the time 
since last merger. Here we summarize how the merger trees are extracted. 

First, our simulation generates output files that contain particle tags and
positions every $100 h^{-1}$~Myr between $z=2$ and $z=0$.
We run a FOF halo finder with linking length parameter $b = 0.2$
on the particle positions to find all the groups containing more than 10 particles. 
Although some of the very small groups are under our halo completeness limit, they have
little effect on our results because we only look at minor or major mergers for which the
mass of the smaller object is above our completeness limit. Thus all mergers for the objects
we study are being counted.
Between successive outputs at times $t = t_n$, we 
find the progenitors at time $t_{n-1}$ for all the halos at $t_n$ by tracing the particle tags, 
which are uniquely assigned to each particle in the beginning of the simulation. A halo $A$ 
is identified as a progenitor of another halo $B$ if $A$ contains at least one particle 
that is also in $B$. For each halo we record the masses of its progenitors, 
their contributed masses, and the number of unbound particles. Then the merger
trees are constructed by linking all the progenitors identified in the previous outputs for
halos above our halo completeness limit at $z=0$. Deriving the mass accretion histories
is straightforwardly accomplished by following the mass of the most massive progenitor back in time.  

To find the time since last merger for any given halo,
we need to define what a `merger' is. There are many 
different ways to define mergers in cosmological simulations. In our analysis
we adopt two definitions: the mass-jump definition, in which a merger is present if there is a 
mass jump in the halo's assembly history; and the mass-ratio definition, which identifies a 
merger if the ratio of contributed masses from the first- and second-ranked progenitors is less
than a certain value \citep{2005APh....24..316C}. To study the variations in cluster observables induced by different types of mergers, we use $1.2$ and $1.33$ as thresholds for the mass-jump definition and
10:1, 5:1, and 3:1 in the mass-ratio definition. For each of the five criteria, the time since 
last merger is found for all clusters at $z=0$ and $z=1$. 

In the later discussions we refer by `merging clusters' at a given lookback time to those 
identified by at least one of the five merger diagnostics in the preceding 3 Gyrs, 
the typical time for clusters to return to virial equilibrium within $R_{500}$ 
\citep{2006MNRAS.373..881P}. The mergers are `major' if the mass jump is larger than 1.2 or 
if the mass ratio is less than 5:1; `minor' mergers, on the other hand, have mass ratios 
between 10:1 and 5:1.


\subsection{Simulated X-ray observations}
\label{Sec:xray}

The observational X-ray temperature of the ICM, $T_X$, usually refers to the spectroscopic
temperature, obtained by fitting a single-temperature model to the integrated cluster 
spectrum. Using simulation-based temperature proxies, such as emission-weighted temperature, 
has been shown to overestimate $T_X$ \citep{Mathiesen:2001,2005ApJ...618L...1R}. 
To directly compare simulated cluster properties with observations, we create mock Chandra 
images for our simulated clusters using the following procedure.

First, energy-dependent surface brightness maps are constructed by projecting 
X-ray emission from all the gas cells associated with each cluster along each of three orthogonal axes. The energy dependence is stored as the third dimension of the map between 0.04 and 10\ keV with
spacing $\Delta$E = 0.0498\ keV. 
With the gas density $\rho$ and temperature $T$ of each cell in each simulation output and an assumed metallicity $Z$, the X-ray emissivity,  $\epsilon_E = \rho^2 \Lambda_E(T, Z, z)$, 
is computed using the single-temperature MEKAL model \citep{Mewe:1985kx,Kaastra:1993yq,Liedahl:1995vn} implemented in the utility XSPEC \citep{1996ASPC..101...17A}.

With these energy-dependent surface brightness maps in hand for each cluster,
we then use MARX\footnote{http://space.mit.edu/ASC/MARX} 
to simulate Chandra X-ray observations. Given the position and
spectrum of a source, MARX can perform ray generation, apply built-in models for 
Chandra's aspect motion and mirror and detector responses, and output photon event files in FITS 
format for analysis with standard observational tools. 
The user-defined source model supported by MARX allows us to generate light rays from 
astronomical sources with arbitrary shape and spectrum. We assign the positions 
and energies of photons based on the probability distribution defined by our 3D surface brightness 
maps. Each cluster is ``observed'' with Chandra's CCD chip ACIS-S 
using an exposure time that ensures 2 million photons are collected, longer than typical
deep observations. The particular long exposure time is chosen to minimize observational uncertainties since we are interested in the intrinsic scatter in cluster observables.
The photon event files thus produced are processed with CIAO to extract the spectra within
apertures of size $R_{500}$ centered on each cluster's surface brightness peak. The spectroscopic 
temperatures $T_X$ are obtained by using XSPEC to fit the spectra in the range 0.5 to 10\ keV,
the range often used by observers \citep[e.g.,][]{2006ApJ...640..691V}.  
We tested the above procedure using a set of isothermal clusters with $\beta$-model density profiles, varying the input temperature from 0.5 to 5\ keV and using two different metallicities, $Z=0$ and $Z=0.3 Z_\odot$. We have verified that for our chosen exposure time, the input temperatures are recovered within $1\sigma$ errors in all cases. For simplicity we assume zero metallicity in the following analysis. 
A detailed description and verification tests of the same X-ray simulator can be found in the appendix of
\cite{2008arXiv0808.0930Z}.

\begin{figure*}[thbp]
\epsscale{1.0}\plottwo{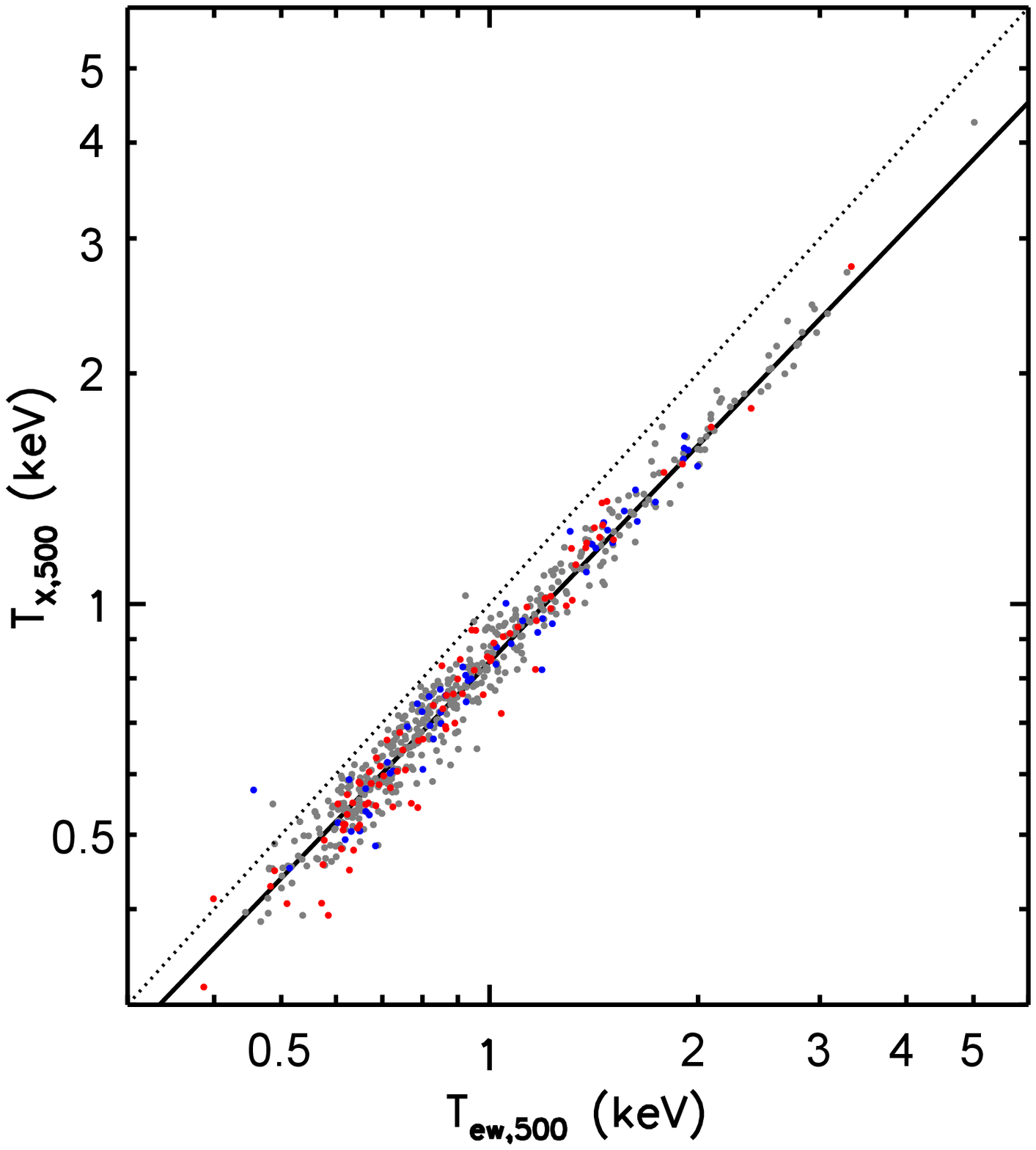}{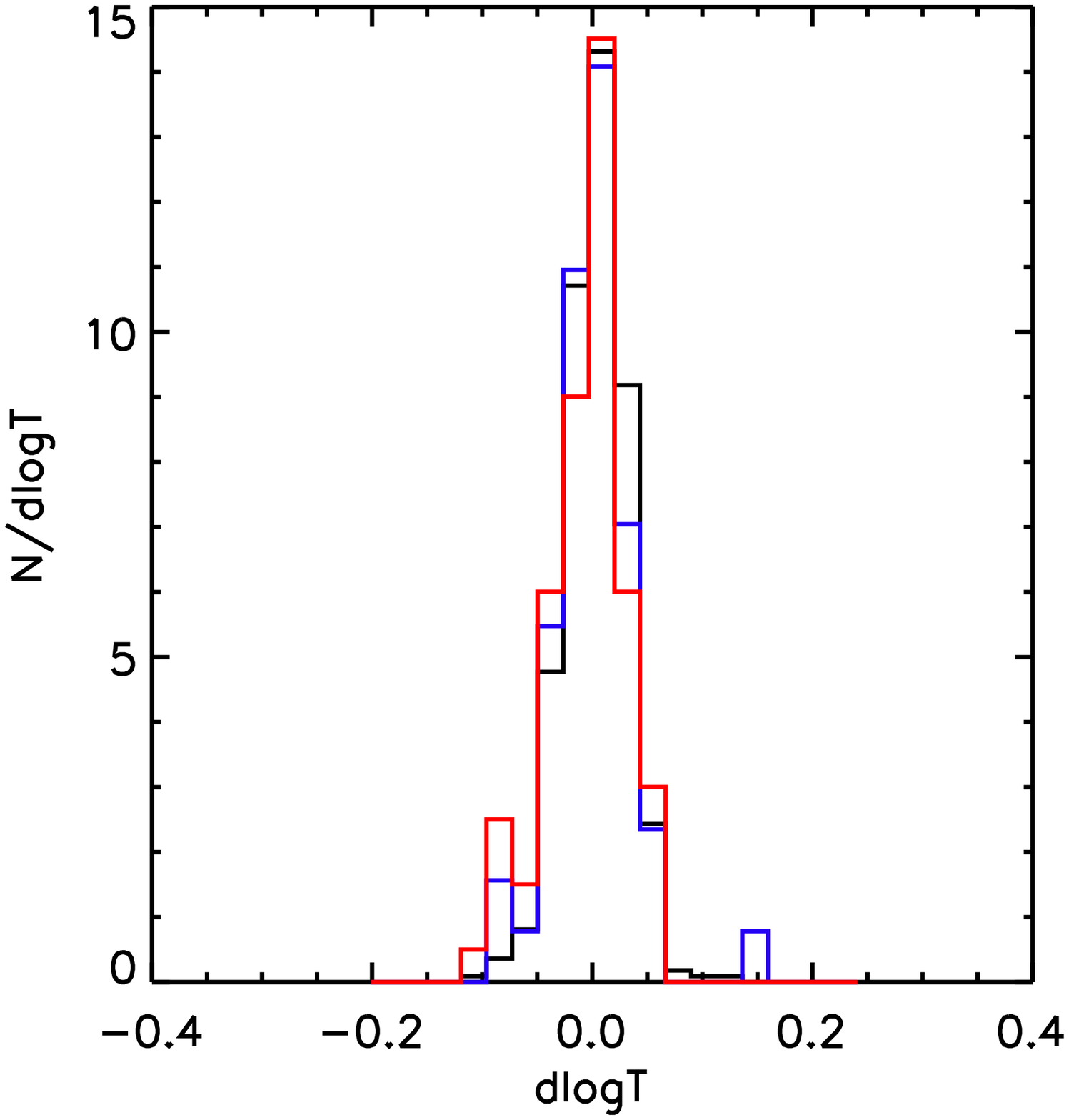}
\caption{{\em Left}: $T_{X,500}$ vs.\ $T_{ew,500}$ for clusters with $M_{500}$ above 
$2\times10^{13} M_\odot$ at $z=0$. The relaxed clusters, minor mergers and major mergers are plotted using grey, blue, and red symbols, respectively. The solid line is the best-fit relation. Comparing to the dotted line with a slope of 1 shows that $T_{ew}$ is biased high with respect to $T_X$.
{\em Right}: Normalized distributions of deviations from the best-fit relation in $T_{X,500}$ vs.\ $T_{ew,500}$ scaling for relaxed clusters (black), minor mergers (blue), and major mergers (red).}
\label{ts}
\end{figure*}

The left panel in Figure \ref{ts} shows $T_X$ versus the emission-weighted temperature $T_{ew}$ measured within $R_{500}$ for clusters with
$M_{500}$ above $2\times10^{13}\ M_\odot$ at $z=0$. We find that $T_{ew}$ is on average biased high with respect to $T_X$ with a fractional bias of $(T_{ew}-T_X)/T_X \sim 23\%$, consistent with previous
findings \citep{Mathiesen:2001,2005ApJ...618L...1R}. 
As pointed out by previous authors, this systematic shift is due to the superposition of cluster gas with different temperatures along the line of sight. To verify that this explains our result, we performed a test for a set of beta-model clusters whose gas is composed of two different temperatures, with varied cool-gas fractions. We found that $T_{ew}$ is higher than $T_X$ in all cases. We can understand this result by expressing these temperature measures in the following way: $T\equiv \int W(T)TdV/\int W(T)dV$, where $W(T)$ is a weighting function. For $T_{ew}$, $W(T)=\Lambda(T) \propto \sqrt{T}$ if dominated by bremsstrahlung emission. Assuming $T_X$ can be approximated by the spectroscopic-like temperature, then $W(T)\sim T^{-3/4}$ \citep{2005ApJ...618L...1R}, that is, $T_X$ tends to weight more on the cool gas in a cluster and thus is systematically lower than $T_{ew}$ in general.

To see whether this systematic shift is dependent on merger types, we also plot in the right panel the normalized distributions of deviations from the best-fit relation for relaxed clusters, minor mergers, and major mergers. Comparing their distributions using the Wilcoxon Rank-Sum test shows that the systematic shift between $T_X$ and $T_{ew}$ is similar for relaxed clusters and minor mergers, but is larger at a statistically significant level for major mergers, based on our merger definitions. This is also in agreement with \cite{Mathiesen:2001}, who found that the difference between $T_X$ and $T_{ew}$ is larger for merging clusters because the spectral fit is largely contributed by X-ray photons from the bright and cool accreted subclumps. 


\subsection{Substructure measures}

High-resolution X-ray images of clusters taken with Chandra and XMM-Newton have revealed 
disturbed ICM structures such as shocks, bubbles, and cold fronts \citep[e.g.][]{2004ApJ...607..800B,2004ApJ...610L..81H}. 
Various substructure measures 
have been used to quantify the irregularity, such as centroid offset \citep{1995ApJ...447....8M}
and power ratios \citep{1995ApJ...452..522B,1996ApJ...458...27B}. In previous work
this irregularity is often 
assumed to be associated with mergers. To examine the effectiveness of the
substructure measures and the adequacy of using them as indicators of dynamical state,
we calculate the centroid offset and power ratios in addition to other theory-based definitions
of cluster mergers in our simulation.

There are many ways to define the centroid offset. For observed clusters, it can be defined as
the variance in the centroids of cluster regions above several surface-brightness isophotes 
\citep[e.g.,][]{2006ApJ...639...64O}. Since the offset is essentially a measure of the distance 
between the surface brightness peak and the cluster centroid, \cite{2007MNRAS.377..317K} 
used a simpler definition for their simulated clusters,
\begin{equation}
w = \frac{ | \vec{R}_{\Sigma,max} - \vec{R}_{\Sigma,cen} | }{R_{500}},
\end{equation}
where $\vec{R}_{\Sigma,max}$ is the position of the surface brightness peak and 
$\vec{R}_{\Sigma,cen}$ is the surface-brightness centroid. We compared the centroid offsets 
calculated using both methods and found that these two definitions give similar results. 
We will use the latter definition hereafter since it is more strongly
correlated with the power ratios for our simulated clusters.

Power ratios are the multipole moments of surface brightness measured within a circular aperture
centered on the cluster's centroid. The moments, $a_m$ and $b_m$ (defined below), are sensitive
to substructures in the surface brightness distribution. This method is motivated by the multipole expansion of the two-dimensional gravitational potential,
\begin{eqnarray}
&\Psi&(R,\phi) = - 2Ga_0 \ln\left( \frac{1}{R} \right) \nonumber \\
&-& 2G \sum^\infty_{m=1} \frac{1}{mR^m} ( a_m \cos m\phi + b_m \sin m\phi )\ . \label{potential}
\end{eqnarray}
The moments $a_m$ and $b_m$ are
\begin{eqnarray}
a_m(R) &=& \int_{R' \leq R} \Sigma( \vec{x}' ) (R')^m \cos m\phi ' d^2 x', \nonumber \\
b_m(R) &=& \int_{R' \leq R} \Sigma( \vec{x}' ) (R')^m \sin m\phi ' d^2 x',
\end{eqnarray}
where $\vec{x}' = (R', \phi ' )$, $R$ is the aperture radius, and $\Sigma$ is the surface mass density,
or surface brightness in the case of X-ray observations. 

The $m$th power $P_m$ is the azimuthal average of the amplitude of $\Psi_m$, 
the $m$th term in the multipole
expansion of the potential given in equation (\ref{potential}),
\begin{equation}
P_m(R) = \frac{1}{2\pi} \int^{2\pi}_0 \Psi_m (R,\phi) \Psi_m (R, \phi) d\phi.
\end{equation}
For $m = 0$ and $m>0$, we have
\begin{eqnarray}
P_0 &=& [a_0 \ln(R)]^2 \nonumber \\
P_m &=& \frac{1}{2m^2 R^{2m}} (a_m^2 + b_m^2),
\end{eqnarray}
respectively. The power ratios are thus $P_m/P_0$, the $m$th power normalized by the flux 
within $R$. 

For each cluster we compute $P_2/P_0$, $P_3/P_0$, and $P_{(1)}/P_0$. The quantities
without parenthetical subscripts are evaluated about the surface brightness
centroid (therefore $P_1$ vanishes by definition). The quadrupole power, $P_2$, 
is related to the degree of flattening or ellipticity. The next odd moment, $P_3$, is sensitive
to unequal bimodal structures. $P_{(1)}/P_0$, which is calculated about the surface brightness
peak, measures gas distribution around the peak and thus is similar to
the centroid offset. Since the power ratios are sensitive to substructures at the scale of the 
aperture radius $R$, each of the power ratios is calculated using three different radii, 
$R_{500}$, $R_{200}$, and 1\ Mpc, for each cluster. 


\section{Results}
\label{Sec:results}

\begin{figure}[thbp]
\begin{center}
\includegraphics[width=0.4\textwidth]{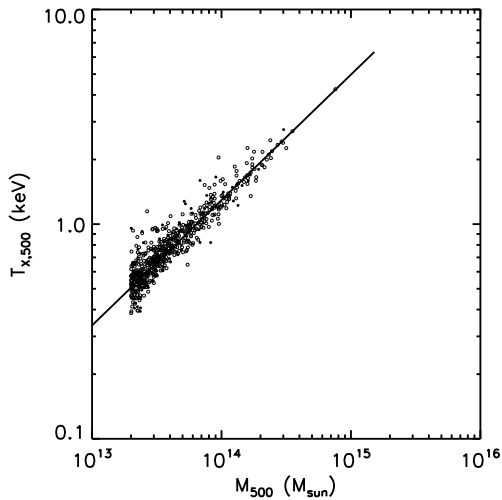}
\caption{$M_{500}$--$T_{X,500}$ relation for clusters with $M_{500}$ above $2\times10^{13} M_\odot$
at $z=0$. Clusters identified by at least one of the five merger diagnostics are plotted using filled circles; relaxed clusters are plotted using open circles. The error bars in $T_X$ ($<0.5\%$) are smaller than the symbol size and are thus omitted. The solid line is the best-fit relation for all clusters.}
\label{mts}
\end{center}
\end{figure}

Figure \ref{mts} shows the $M_{500}$--$T_{X,500}$ relation for clusters with $M_{500}$ above $2\times10^{13} M_\odot$ at $z=0$. All quantities are viewed along the $x$-direction in the simulation box. 
The best-fit relations for all clusters as well as different subgroups are given in Table \ref{mts_fit}. 
Note that the best-fit slopes for relaxed and merging clusters are different. 
The difference in the slopes implies that there is variation in the slopes derived from observed clusters using different selected subsamples. This systematic uncertainty in the slope, which is $\sim 5\%$ by comparing the slopes of the merging and relaxed clusters in Table \ref{mts_fit}, will contribute to the uncertainties in cosmological parameters.
For example, it will translate into a systematic uncertainty of $\sim 3\%$ in $\sigma_8$ when using observed cluster samples including and excluding merging clusters. This is relatively small compared to the current level of uncertainty ($\sim 10\%$) in determining the cosmological parameters, but as the constraints are improved to a few percent in the future, one has to keep it in mind when comparing results using different cluster selection criteria.

While we are able to reproduce the self-similar $M_{200}$--$T_{ew}$ scaling relation \citep[e.g.,][]{1998ApJ...495...80B}, the slope and normalization of the $M_{500}$--$T_{X,500}$ relation are both smaller compared to observed values. This deviation is mainly due to missing baryonic physics and the use of different mass estimates \citep[e.g.,][]{2004MNRAS.348.1078B, 2007ApJ...668....1N}. 
Matching the slope and normalization of the scaling relations with observed values is a subject of great interest on its own and will be investigated in our future papers when we incorporate more realistic models for the baryonic component. For the purpose of this paper, we will focus on the scatter in these relations and the relative importance of clusters of different merger histories. 

\begin{table}[thdp]
\caption{Best-fit parameters in the mass-temperature relations, $\log(T_{X,500}) = a+b\log(M_{500})$, 
for clusters at $z=0$. Errors in the parameters are the $1\sigma$ errors.}
\begin{center}
\begin{tabular}{lrcc}
\hline
\hline
Subgroup & Count & $a$ & $b$ \\
\hline
All & 619 & -8.067 $\pm$ 0.004 & 0.5842 $\pm$ 0.0003\\
Relaxed & 478 & -7.981 $\pm$ 0.004 & 0.5780 $\pm$ 0.0003\\ 
Merging & 141 & -8.409 $\pm$ 0.008 & 0.6093 $\pm$ 0.0006\\
\ \ Minor & 55 & -8.379 $\pm$ 0.013 & 0.6068 $\pm$ 0.0009\\
\ \ Major & 86 & -8.050 $\pm$ 0.011 & 0.6163 $\pm$ 0.0008\\
\hline
\hline
\end{tabular}
\end{center}
\label{mts_fit}
\end{table}


\subsection{Distribution of intrinsic scatter}
\label{Sec:scatter distribution}

\begin{figure*}[thbp]
\epsscale{1.0}\plottwo{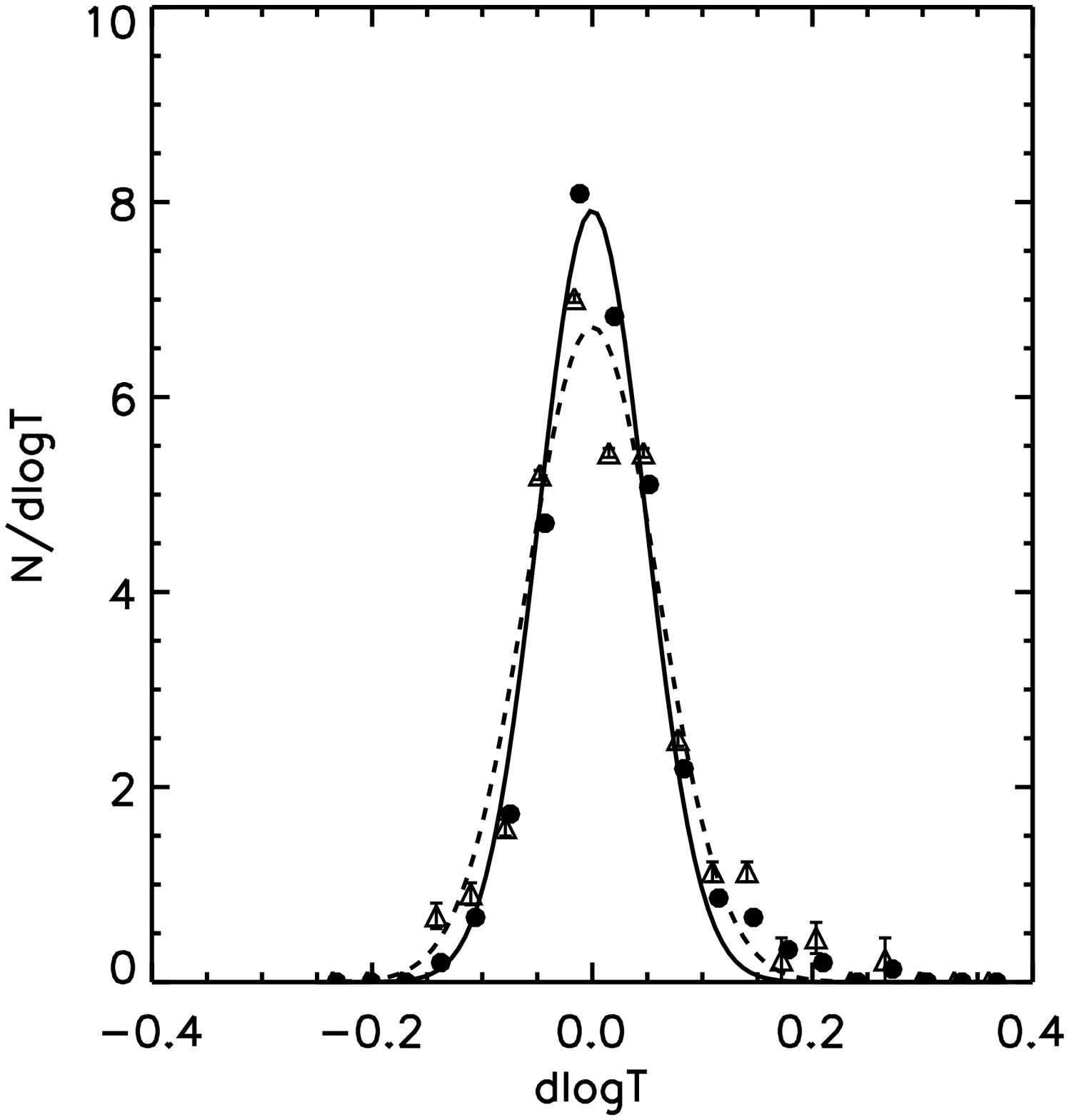}{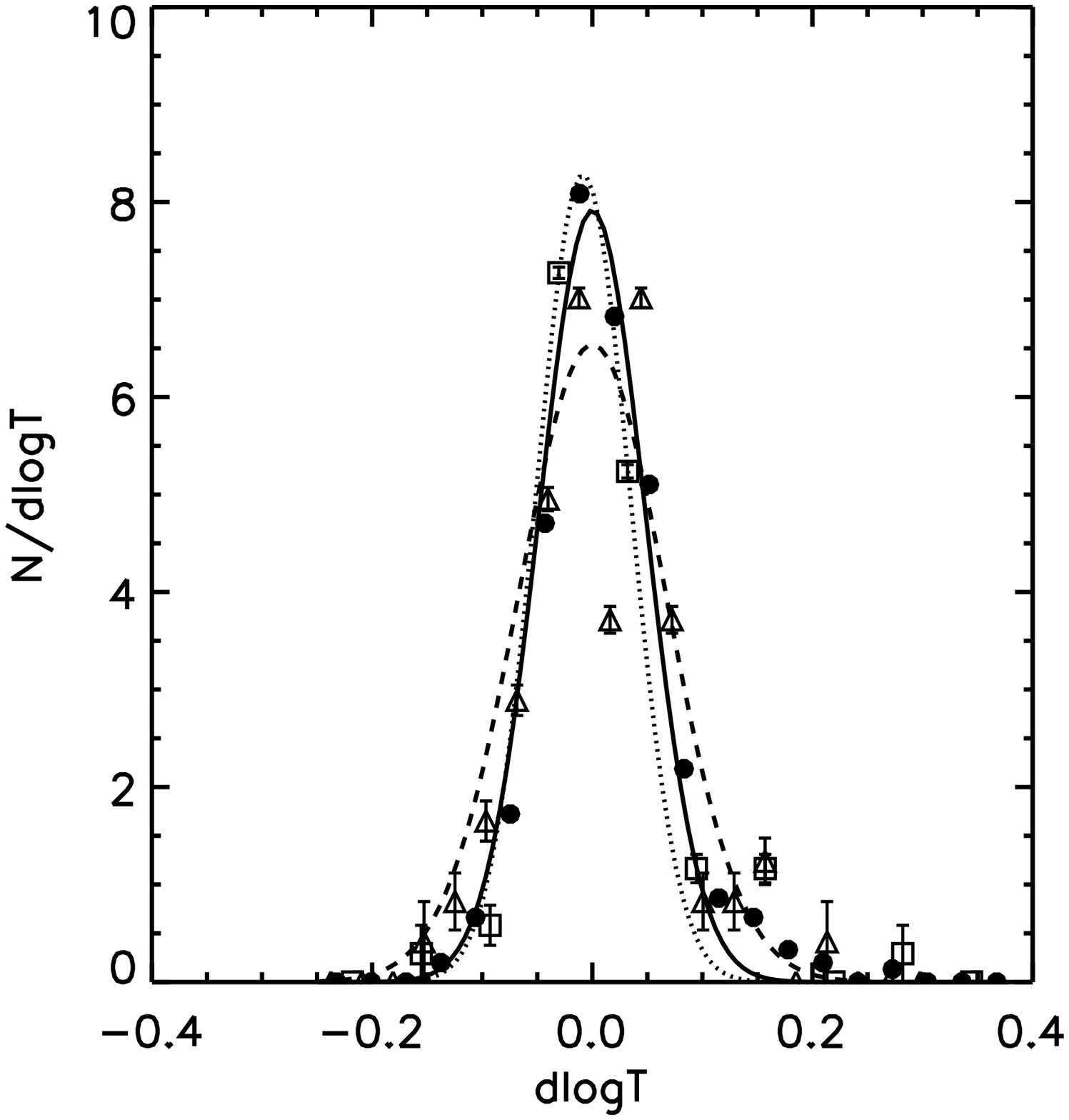}
\caption{{\em Left}: Normalized distribution of log scatter in the $M_{500}$--$T_{X,500}$ relation
at $z=0$ for merging and relaxed clusters. The curves are the best-fit lognormal distributions. Merging clusters are plotted using open triangles and dashed lines; relaxed clusters are plotted using filled circles and solid lines.
{\em Right}: Normalized distribution of log scatter and the best-fit curves for minor mergers (open squares and dotted lines), major mergers (open traiangles and dashed lines) and 
relaxed clusters (filled circles and solid lines).}
\label{mtdev_dist_z0}
\end{figure*}

Figure~\ref{mtdev_dist_z0} shows the normalized distribution of the logarithmic deviations 
of temperature from the best-fit $M_{500}$--$T_{X,500}$ relation at $z = 0$. 
The RMS scatter for our whole sample is $6.10\%$, which is smaller than the values obtained by simulations with cooling and heating, such as $13.6\%$ in \cite{2007ApJ...668....1N} and $20\%$ in \cite{2006ApJ...639...64O}, who also used the spectroscopic temperature and true mass. It is difficult to directly compare with observed values because the observationally-estimated intrinsic scatter is dependent on how the measurement errors are determined and thus displays a wide range in the literature, from $3.9\%$ in \citep{2005A&A...441..893A} to $17\%$ in \citep{2006ApJ...639...64O}.

In the left panel of Figure~\ref{mtdev_dist_z0} we compare the distributions of merging (red) and relaxed (black) clusters, while the right panel plots for minor and major mergers individually.
The two distributions on the left display no apparent difference in their mean values. The standard deviation for merging clusters appears to be larger than 
that of the relaxed clusters. To test the hypothesis that the merging and relaxed distributions differ,
we performed the Wilcoxon Rank-Sum (R-S) test and the F-variance (F-V) test
to see whether these two populations have significantly different
mean values or variances, respectively. A small value ($<0.05$ for a significance level of $5\%$) 
returned by the tests is often adopted to indicate a significant difference between these two populations. 
We performed the tests on different merging subgroups and summarize the results in Table~\ref{sig_test}.  

The R-S test results show that the mean values do not differ significantly among all populations, which means that the intrinsic scatter is unbiased for merging and relaxed clusters at $z=0$. The F-V tests for all mergers and minor mergers show that their standard deviations, or the amount of scatter, are significantly larger than that of the relaxed ones. For clusters at $z=1$, there is also no bias between merging and relaxed populations. The amount of scatter for merging clusters also tends to be greater than relaxed clusters, although only major mergers show a significant result. We will discuss the possible reasons for this trend in the next two sections.

Since the form of scatter can affect the observed scaling relations, we also test the
Gaussianity of the distribution of scatter in log space by fitting it with a Gaussian curve and 
using the Kolmogorov-Smirnov (K-S) test to see if the two distributions differ significantly.
Again, a small value represents a significant deviation from the Gaussian distribution.
The test results show that the distributions of scatter for all populations but minor mergers
differ from a lognormal distribution at a significant level. There is an even stronger signature of 
deviation from lognormal at $z=1$. 
We note that the deviation of the scatter from a lognormal distribution may affect results from some of the 
self-calibration studies assuming a lognormal distribution of scatter \citep[e.g.,][]{2005PhRvD..72d3006L} and should be taken into account to correctly interpret the obtained constraints on the cosmological parameters. 

\begin{table*}[thdp]
\caption{Significance tests on the distribution of scatter for different populations at $z=0$ and $z=1$. 
The R-S and F-V test results for each subgroup are relative to the relaxed clusters.}
\begin{center}
\begin{tabular}{cccccccc}
\hline
\hline
Subgroup & $z$ & $N$ & Mean &  R-S Test & $\sigma_{rms} (\%)$ & F-V test & K-S test \\
\hline
Relaxed & 0 & 478 & $1.31\times10^{-2}$ & - & 5.87 & - & $2.5\times10^{-2}$ \\ 
Merging & 0 & 141 & $8.67\times10^{-3}$ & 0.134 & 6.90 & 0.014 & $2.5\times10^{-2}$ \\ 
Minor & 0 & 55 & $1.11\times10^{-2}$ & 0.146 & 7.21 & 0.029 & $2.0\times10^{-1} $ \\ 
Major & 0 & 86 & $7.09\times10^{-3}$ & 0.253 & 6.72 & 0.086 & $2.5\times10^{-2}$ \\ 
\hline
Relaxed & 1 & 102 & $7.14\times10^{-3}$ & - & 4.85 & - & $2.6\times10^{-4}$ \\ 
Merging & 1 & 121 & $8.72\times10^{-3}$ & 0.493 & 5.71 & 0.091 & $6.5\times10^{-3}$ \\
Minor & 1 & 46 & $1.41\times10^{-2}$ & 0.230 & 5.25 & 0.504 & $8.0\times10^{-2}$ \\
Major & 1 & 75 & $5.45\times10^{-3}$ & 0.290 & 5.98 & 0.049 & $6.5\times10^{-3}$ \\
\hline
\hline
\end{tabular}
\end{center}
\label{sig_test}
\end{table*}


\subsection{Intrinsic scatter vs. halo concentration}
\label{Sec:concentration}

In the following two sections we will investigate the physical origin of the $M_{500}$--$T_{X,500}$ scatter by correlating it with cluster properties that are related to how the clusters are formed. In particular,
we first show that the intrinsic scatter depends strongly on halo concentration. Then we discuss the contribution of scatter from recent merging events in the following section. 

\begin{figure*}[htbp]
\begin{center}
\epsscale{1.0}\plottwo{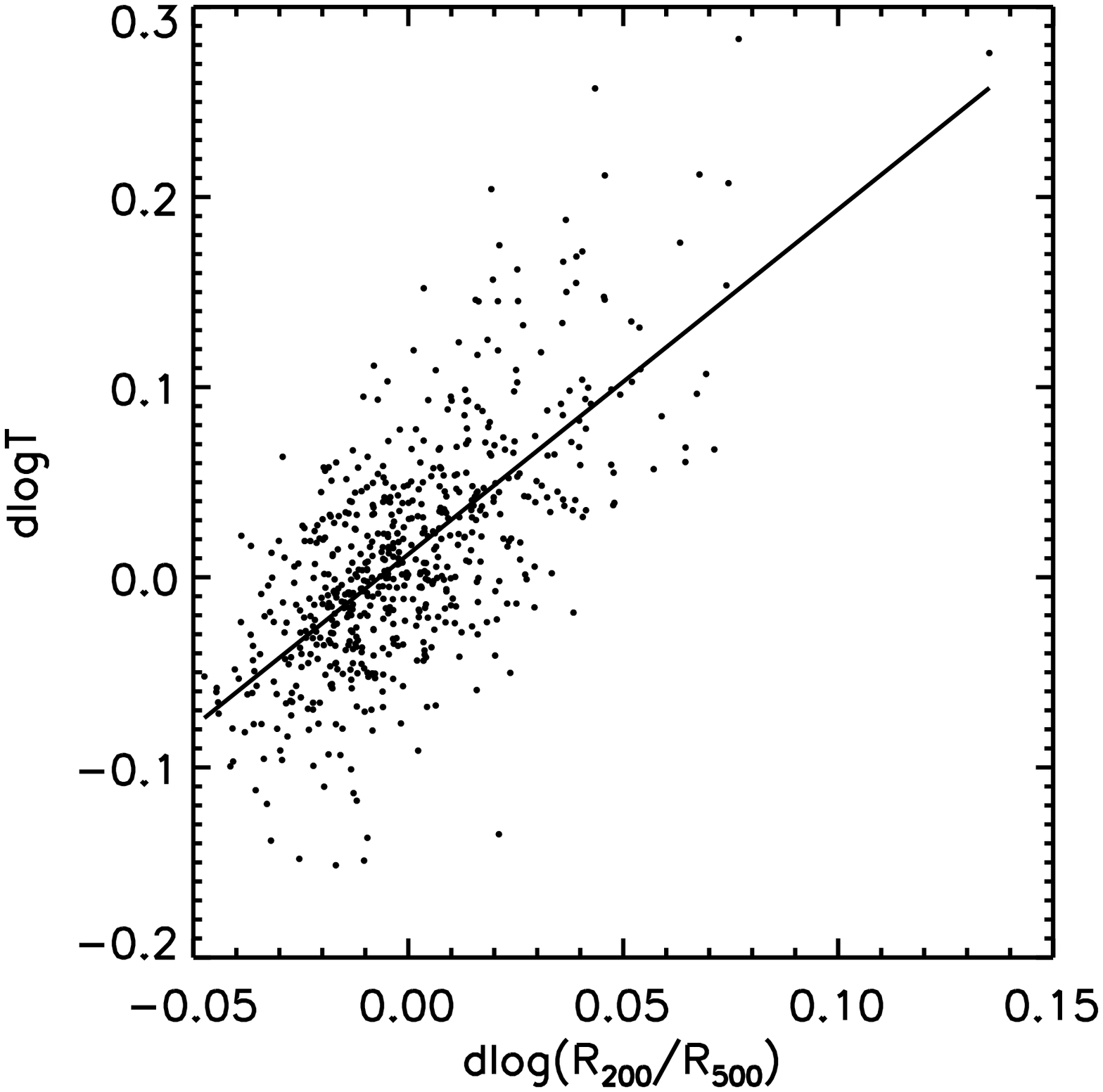}{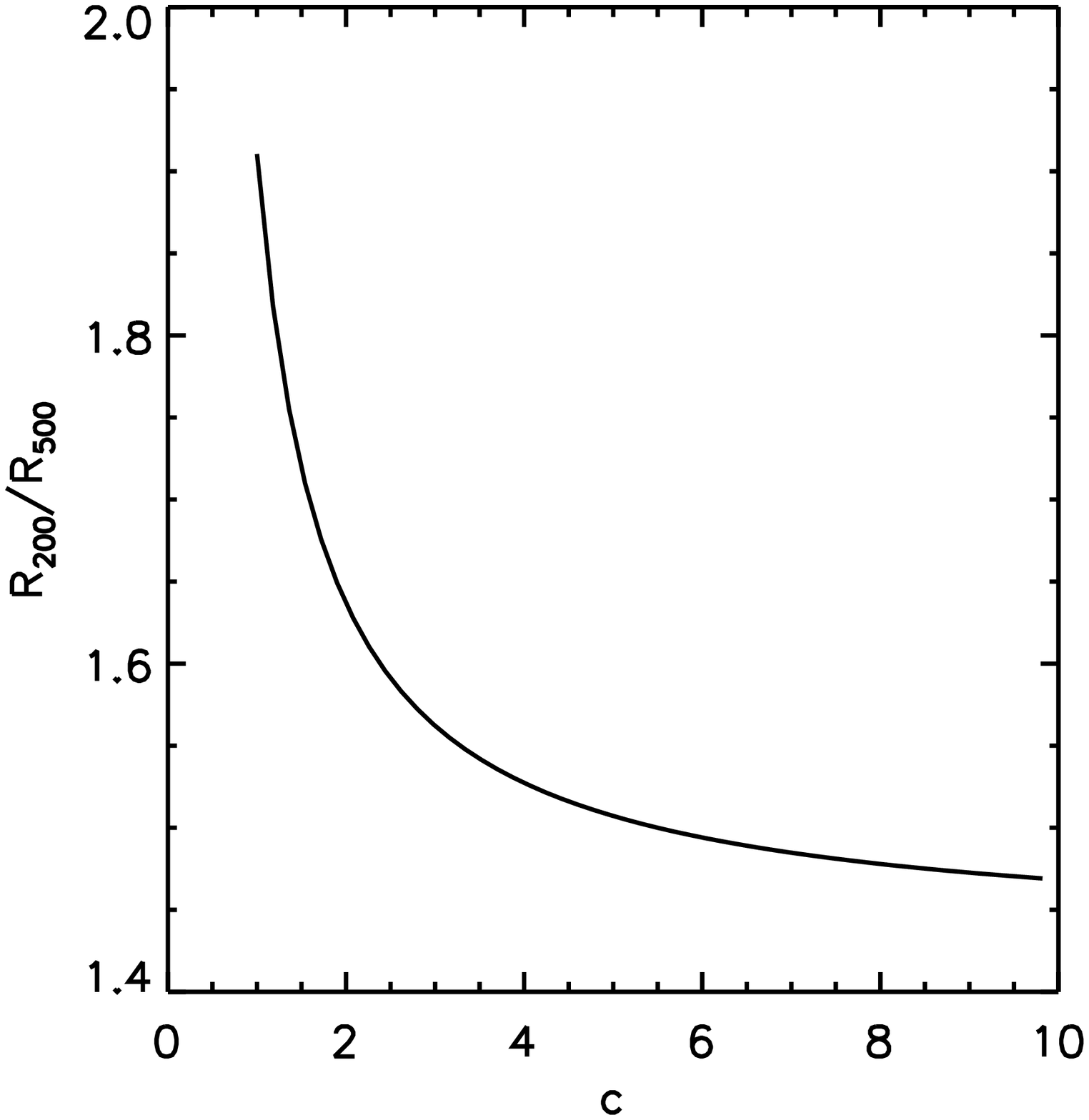}
\caption{{\em Left:} $M_{500}$--$T_{X,500}$ scatter versus $M_{500}$--$R_{200}/R_{500}$ scatter. There is a significant positive correlation, with correlation coefficient 0.64. The solid line is the best-fit relation $d\log T_X = 1.81\times d\log(R_{200}/R_{500})$. {\em Right:} Relation between $R_{200}/R_{500}$ and the halo concentration parameter, $c\equiv R_{200}/R_s$, for an NFW profile. }
\label{mtsdev_c}
\end{center}
\end{figure*}

Figure \ref{mtsdev_c} shows a strong positive correlation between the scatter in 
the $M_{500}$--$T_{X,500}$ relation and the scatter in the $M_{500}$--$(R_{200}/R_{500})$ relation.
The correlation coefficient is 0.64, with a probability of zero, given by the Spearman Rank-Order Correlation test (\citeauthor{1992NumericalRecipe}\ 1992, \S14.6; probability of one means no correlation).  
To ensure that this result is not biased by the lower-mass clusters whose $R_{500}$ values are close to the resolution of the simulation, we raised the mass threshold to $M_{500} \geqslant 10^{14} M_\odot$ and repeated the analyses for these well-resolved systems. We found that for the 67 selected massive clusters, the correlation still holds, with Spearman correlation coefficient 0.36 and probability 0.003.
Note that we correlate with $d\log(R_{200}/R_{500})$ instead of the raw value of $R_{200}/R_{500}$ because the latter is a function of cluster mass. By doing so we exclude the effect of
different cluster masses, focusing on the variation in halo concentrations.
$R_{200}/R_{500}$ is a monotonically decreasing function of the halo concentration parameter, usually defined as $c\equiv R_{200}/R_s$, where $R_s$ is the scale radius of a cluster. 
Therefore, for clusters with similar masses, more concentrated clusters tend to lie under the mean mass-temperature relation, while the puffier clusters tend to scatter high. 

The relation between $R_{200}/R_{500}$ and $c$ is less obvious, 
so we derive their relation assuming an NFW profile (Navarro, Frenk, \& White \citeyear{1995MNRAS.275..720N, 1996ApJ...462..563N}, hereafter NFW) in the following. 
For a cluster that has an NFW density profile, the mass enclosed within a normalized radius of $x\equiv R/R_s$ is
\begin{eqnarray}
M(<x) &=& 4\pi \rho_s R_s^3 \left[ \ln (1+x) - \frac{x}{1+x} \right] \nonumber \\
&\equiv& 4\pi \rho_s R_s^3 f(x), \label{MNFW}
\end{eqnarray}
where $\rho_s$ is the density at the scale radius $R_s$. Also, the mass in the spherical overdensity definition is 
\begin{equation}
M_\Delta = \frac{4\pi}{3} \Delta \rho_{crit} (x_\Delta R_s)^3, \label{Mso}
\end{equation}
where $\Delta$ is the overdensity and $\rho_{crit}$ is the critical density of the universe. Equating equation (\ref{MNFW}) and (\ref{Mso}) gives
\begin{equation}
\Delta \frac{x_\Delta^3}{f(x_\Delta)} = 3 \frac{\rho_s}{\rho_{crit}},
\end{equation}
which is a constant for the cluster under consideration. Therefore, the relation between $x_{500}$ and $x_{200}$ (or $c$, recall $c\equiv R_{200}/R_s)$ is
\begin{equation}
\frac{x_{500}^3}{f(x_{500})} = \frac{2}{5} \frac{c^3}{f(c)}.
\end{equation}
We can use this relation to numerically solve for $x_{500}$ as a function of $c$, and then the relation between $R_{200}/R_{500}$ and $c$ is simply
\begin{equation}
\frac{R_{200}}{R_{500}} = \frac{c}{x_{500}(c)}.
\end{equation}
This relation is plotted in the right panel of Figure \ref{mtsdev_c}. We choose to use the parameter $R_{200}/R_{500}$ instead of the original halo concentration parameter $c$ because it has two advantages. The first is to avoid introducing the uncertainty in fitting an NFW profile, especially for less massive clusters, since the fitting is very sensitive to the resolution in the central region of the cluster. Moreover, our analyses involve not only relaxed clusters but also merging ones, for which $R_{200}/R_{500}$ is actually more well-defined than $c$.

One can understand why the correlation between the $M_{500}$--$T_{X,500}$ scatter and the halo concentration exists using the virial theorem. 
Consider the simplest case for an isolated system: $2T+W=0$, where $T$
and $W$ are the total kinetic and gravitational binding energy of the system, respectively. Then 
in general,   
\begin{equation}
\frac{k_B T_{vir}}{\mu m_p} \propto \frac{GM_{vir}}{R_{vir}},
\end{equation}
where $k_B$ is the Boltzmann constant, $\mu$ is the mean molecular mass 
of the gas, $m_p$ is the mass of a proton, 
and $T_{vir}$, $M_{vir}$, and $R_{vir}$ are the virial gas temperature, mass, and radius of the system. Since the relations between the virial quantities and the quantities in the overdensity definition depend on individual cluster profiles, or halo concentrations, 
the $M_\Delta$--$T_\Delta$ relation derived from above would have 
a normalization which is a function of concentration.
This is why we expect the scatter in the $M_{500}$--$T_{X,500}$ relation to 
correlate with the concentration parameter.

\begin{figure*}[thbp]
\begin{center}
\epsscale{1.0}\plottwo{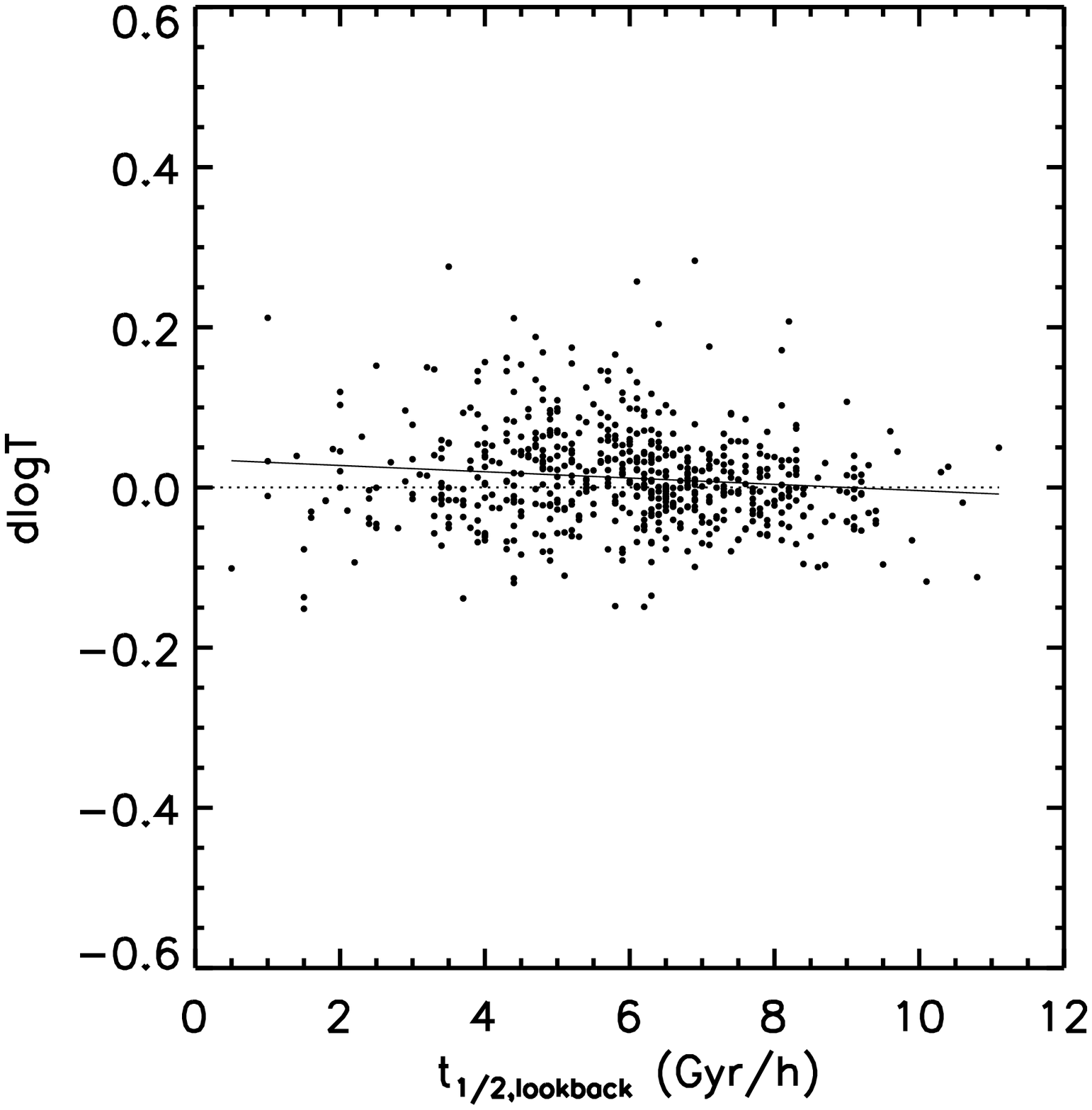}{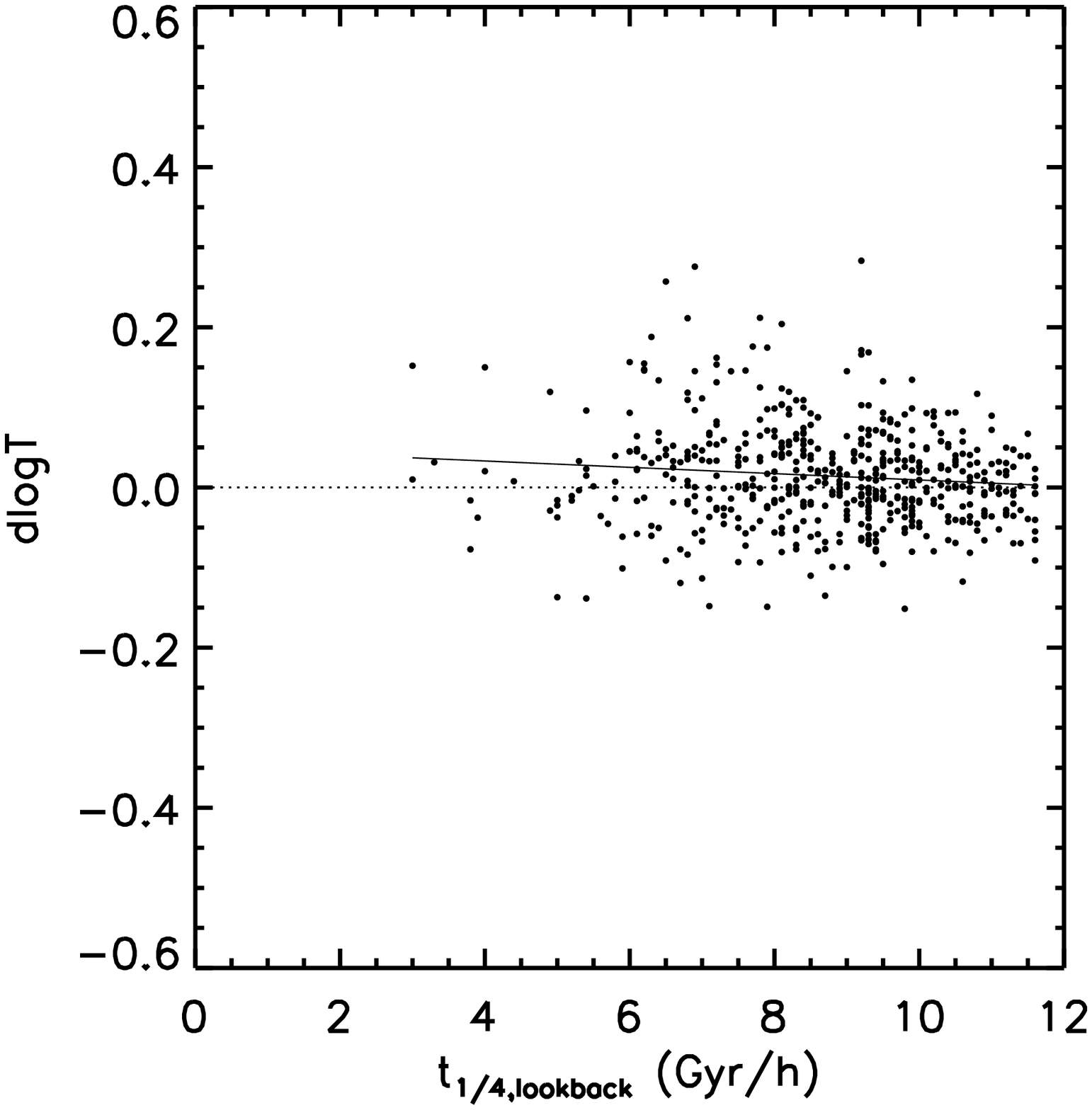}
\caption{{\em Left:} $M_{500}$--$T_{X,500}$ scatter correlated with cluster formation time, defined as the time when a cluster first obtained half of its final mass. The Spearman correlation coefficient is -0.14, with a probability of $1.95\times10^{-4}$ (probability of one means no correlation). {\em Right:} Same plot with x-axis being the time a cluster reached 1/4 of its final mass. The correlation coefficient is -0.11, with a probability of $9.4\times10^{-3}$. Both plots show that clusters that are formed at earlier/later times tend to scatter low/high.}
\label{mts_tform}
\end{center}
\end{figure*}

\begin{figure}[thbp]
\begin{center}
\includegraphics[width=0.4\textwidth]{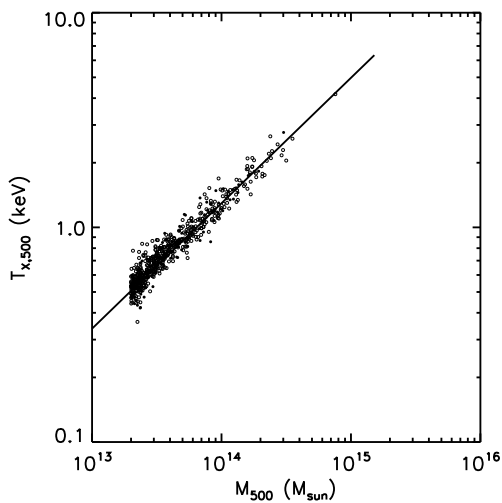}
\caption{$M_{500}$--$T_{X,500}$ relation corrected for halo concentrations. Notations are the same as in Figure \ref{mts}.}
\label{mts_correctc}
\end{center}
\end{figure}

Halo concentrations have been shown to be related to the epoch at which the halo formed 
(NFW \citeyear{1997ApJ...490..493N}; \citeauthor{2002ApJ...568...52W} 2002; \citeauthor{2003MNRAS.339...12Z} 2003; \citeauthor{2007MNRAS.381.1450N} 2007). Since we have the mass assembly histories of all the simulated clusters, it is straightforward to derive the cluster formation
time based on the definition that a cluster ``forms'' when it first exceeds a certain fraction of its final mass.
The commonly-adopted thresholds include 10\%, 25\%, 50\%, and 70\%. Since more concentrated halos tend to form at higher redshift, a negative correlation between the $M_{500}$--$T_{X,500}$ scatter and the formation redshift is expected. Indeed we find significant negative correlations with the time when the cluster first reached one half, $t_{1/2}$, and one quarter, $t_{1/4}$, of its final mass (see Figure \ref{mts_tform}). These correlations are not as tight as the one with halo concentrations, probably due to the fact that the correlation between the halo concentration and the cluster formation time itself has a very large scatter, and also that the variation in halo concentrations cannot be fully accounted for by the variation in cluster formation time \citep{2007MNRAS.381.1450N}. But the significance of these correlations with cluster formation times supports our finding that the scatter correlates with halo concentrations. Therefore, we can say that cluster assembly histories leave imprints on the shapes of clusters at the present day that help to determine clusters' positions on the mass-observable scaling relation.

The strong correlation in Figure \ref{mtsdev_c} suggests that the variation in halo concentrations contributes to a significant amount of the intrinsic scatter in the $M_{500}$--$T_{X,500}$ relation. It also implies that when we are provided 
the best-fit relation, $d\log T_X = 1.81\times d\log(R_{200}/R_{500})$, it is possible to
use $d\log(R_{200}/R_{500})$ as a third parameter to normalize $T_X$ in the $M_{500}$--$T_{X,500}$
relation, and thus reduce the scatter. After removing the effect of halo concentrations, we find that 
the RMS scatter decreases from 6.10\% to 4.49\%, a reduction of $\sim 26\%$ of its original value. The corrected $M_{500}$--$T_{X,500}$ relation is shown in Figure \ref{mts_correctc}. 

\begin{figure*}[htbp]
\begin{center}
\epsscale{1.0}\plottwo{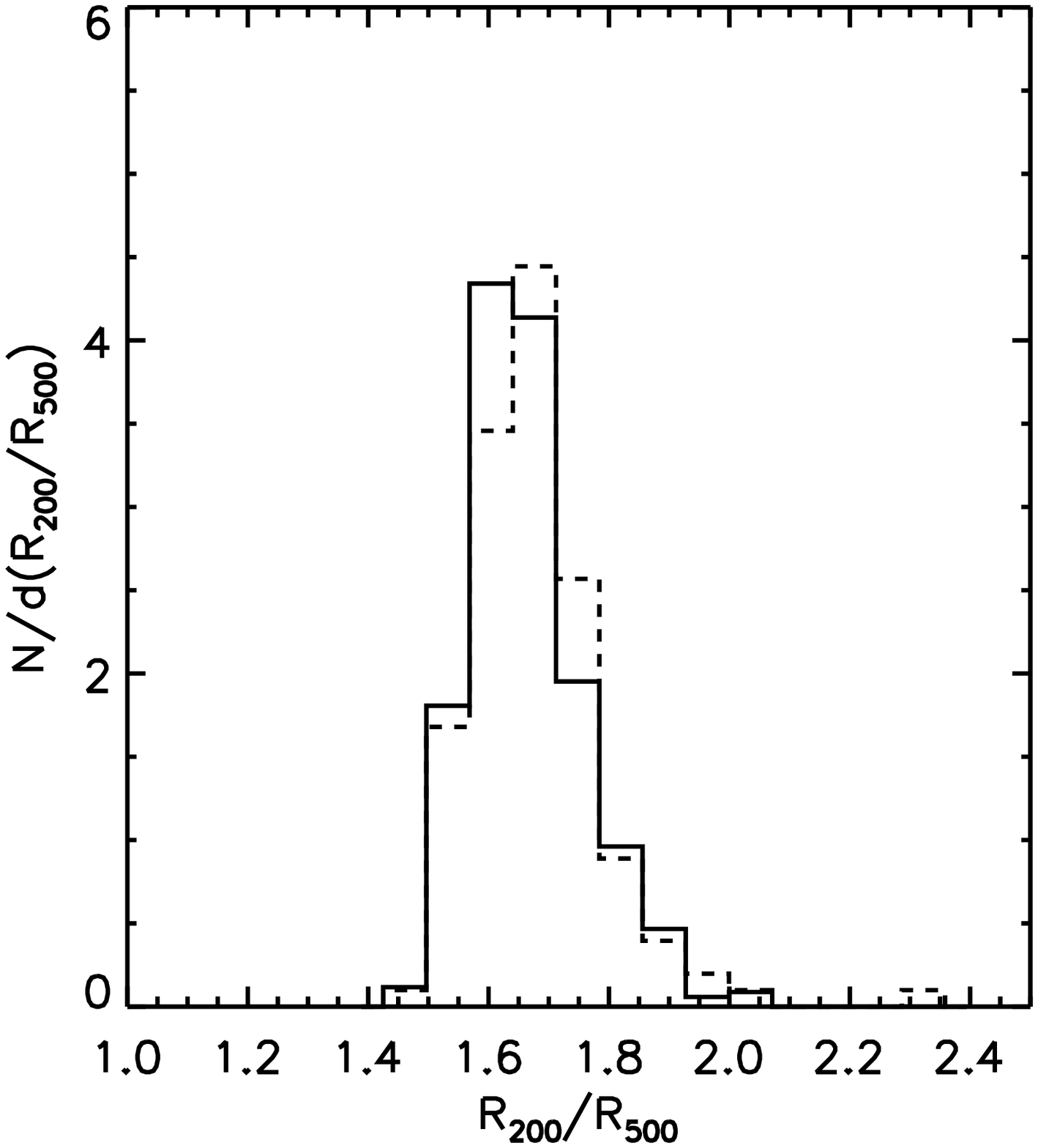}{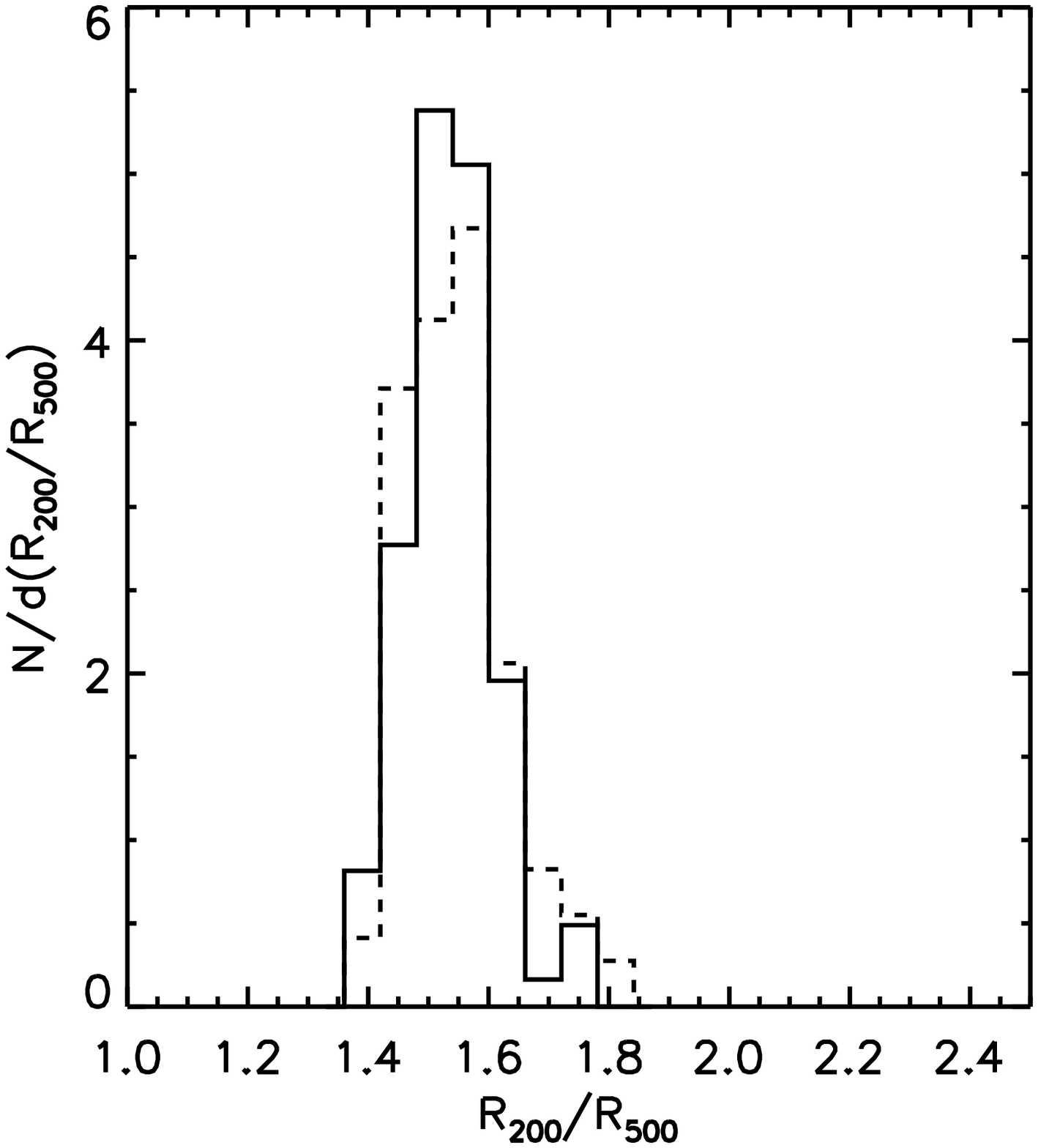}
\caption{Distribution of $R_{200}/R_{500}$ for relaxed (solid) and merging (dashed) clusters at $z=0$ (left panel) and $z=1$ (right panel). According to the F-V test, merging clusters have larger dispersions than relaxed clusters for both $z=0$ and $z=1$ datasets, with significance 0.0044 and 0.0367, respectively.}
\label{c_dist}
\end{center}
\end{figure*}

We can explain the trend found in the previous section using this correlation too. 
Figure \ref{c_dist} shows the distributions of halo concentration for merging and relaxed clusters at $z=0$ and $z=1$. For both redshifts, the distribution of halo concentration for merging clusters is more dispersed than for relaxed clusters, as also found in \cite{2007MNRAS.381.1450N}. If the variation in halo concentrations is important to the $M_{500}$--$T_{X,500}$ scatter, as the above correlation suggests, then it is reasonable that merging clusters have a greater amount of scatter than relaxed ones. One may try to relate this trend to the dynamical state of clusters because the temperature excursions during mergers could also drive the scatter. However, in the next section we will show that this is not the case.  

In summary, we have found that the scatter in the $M_{500}$--$T_{X,500}$ relation partly originates from the variation in halo concentrations, with more concentrated or early-formed clusters lying below the mean (they are cooler), while puffier clusters that are formed recently tend to be hotter than clusters with similar masses. Using the strong correlation between the scatter and halo concentrations, the scatter can be greatly reduced to get a much tighter relation to be used for cosmology. The correlation can also explain the trend seen in the previous section that merging clusters have a greater amount of scatter than relaxed ones.
Note that our simulation adopted the value of $\sigma_8=0.74$ from WMAP3 results, which is smaller than $\sigma_8=0.796$ from WMAP5. Although choosing a smaller $\sigma_8$ would decrease the average concentration of clusters with a fixed mass \citep[e.g.][]{2008MNRAS.390L..64D}, it is the {\it variation} of concentration that correlates with the scatter. Therefore the correlation should still hold if a higher $\sigma_8$ is used.


\subsection{Intrinsic scatter vs. recent mergers}

The other possible origin of the scatter in our simulation is the departure from hydrostatic equilibrium due to cluster mergers. 
Cluster mergers are among the most energetic events in the universe. When two clusters merge, their gas is compressed and heated by merger shocks. This effect can boost the luminosity and temperature of the cluster a few times higher than its pre-merger value, as found in ideal merger simulations \citep{2001ApJ...561..621R, 2007MNRAS.380..437P}. Therefore, our aim is to investigate how merger events statistically influence the cluster scaling relations. 

In order to see how merging events influence the observable quantities of clusters, we correlate their mass-temperature scatter with their dynamical state. Two different methods are used to quantify the dynamical state. The first is based on the actual cluster merger histories, where we use the time since last merger as an indicator. The second one is motivated by observations that unrelaxed clusters often have more substructures than relaxed clusters. We discuss results using both methods in the following. 

\begin{figure*}[htbp]
\begin{center}
\epsscale{1.0}\plottwo{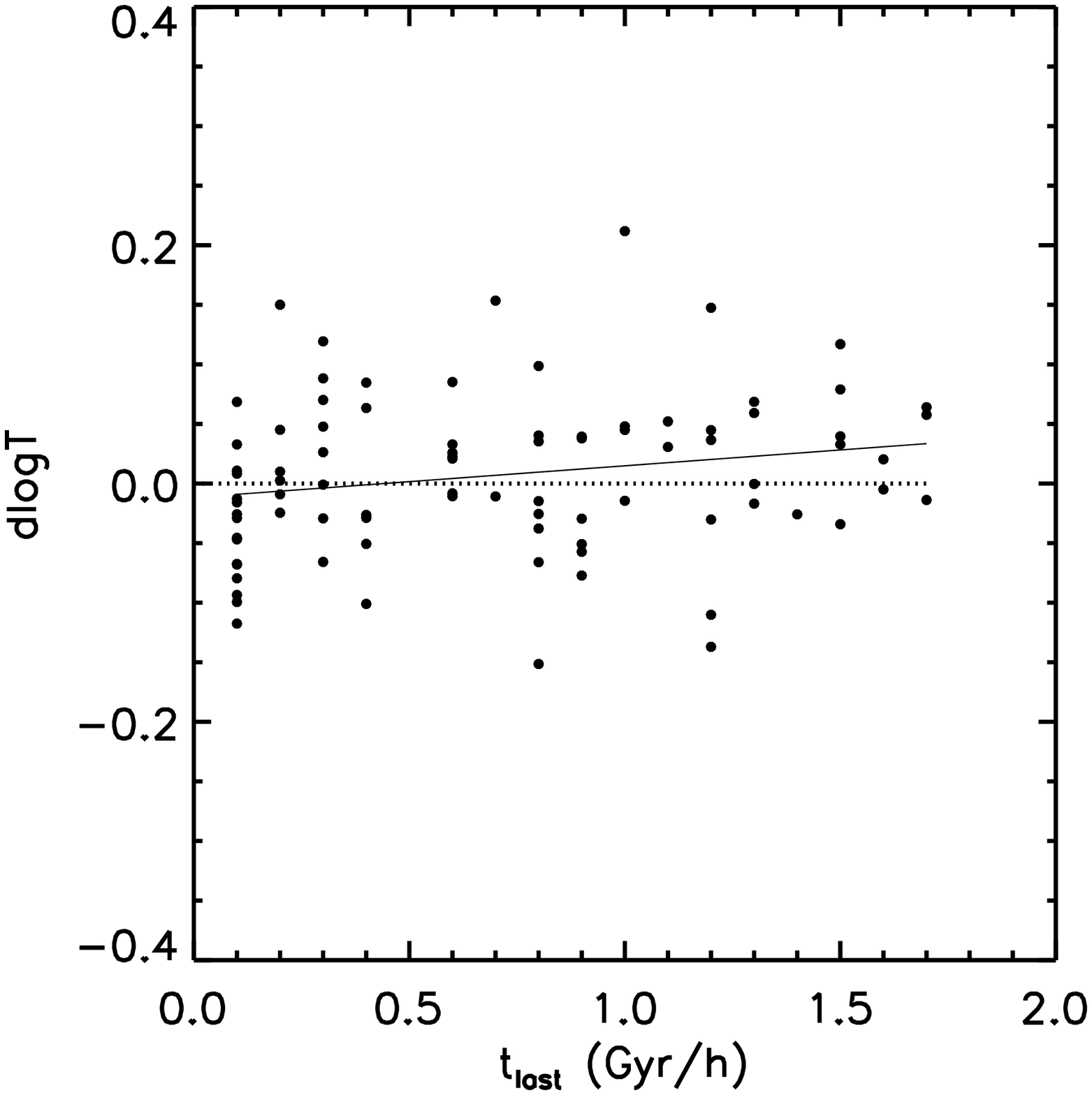}{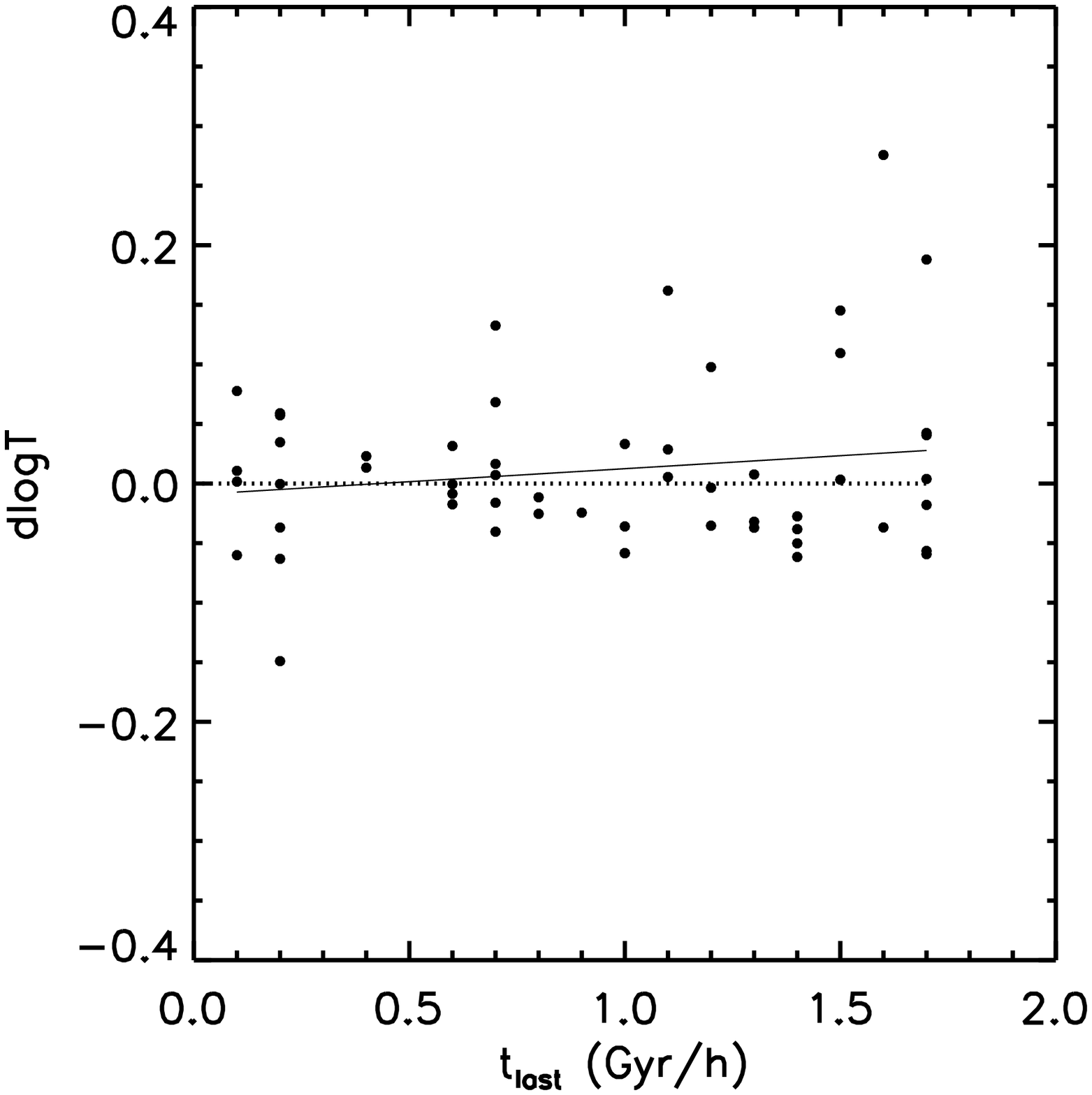}
\caption{$M_{500}$--$T_{X,500}$ scatter versus the time since last merger for major mergers (left panel) and minor mergers (right panel) at $z=0$. The Spearman correlation coefficients are 0.24 and 0.008, with probabilities of no correlation being 0.02 and 0.95 for major and minor mergers, respectively.}
\label{dt_tlast}
\end{center}
\end{figure*}

\begin{figure*}[pthb]
\begin{center}
\includegraphics[height=7.25in]{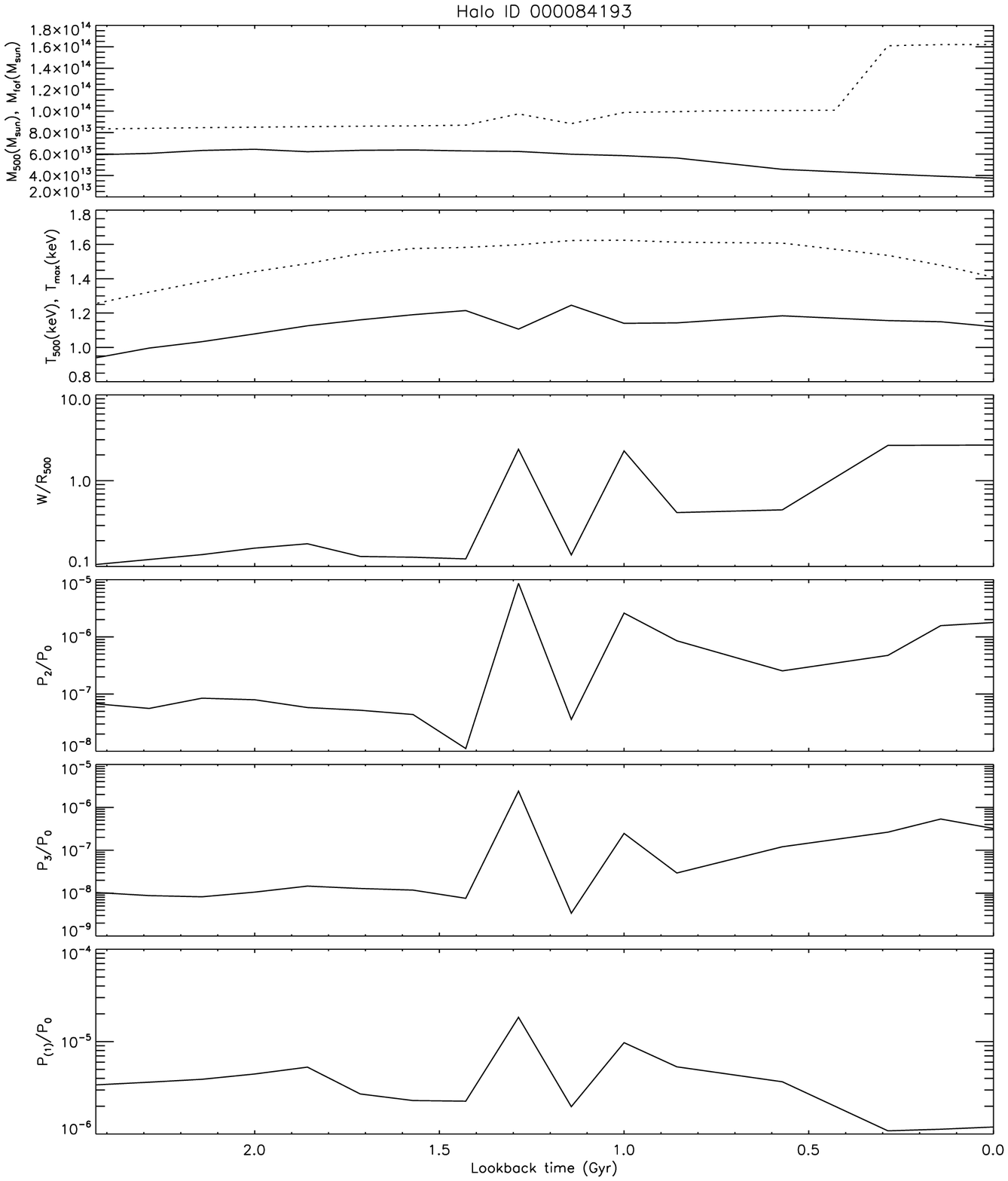}
\caption{Example time history of a cluster that undergoes a minor merger. The solid (dashed) line 
is $M_{500}$ ($M_{fof}$) in the first panel, $T_{500}$ ($T_{max}$) in the second panel. The
other panels show the evolution of the substructure measures.}
\label{hist_minor}
\end{center}
\end{figure*}

First we correlate the $M_{500}$--$T_{X,500}$ scatter with the time since last merger, $t_{last}$ (thus clusters that just underwent mergers would have $t_{last} = 0$). As shown in Figure \ref{dt_tlast}, there is a trend for more recently merged clusters to lie below the mean relation, but only the correlation for major mergers has a high probability. Note that although the scatter here is uncorrected, this correlation is not due to the effect of halo concentrations because halo concentrations work in the opposite direction, as described in \S~\ref{Sec:concentration}. Therefore, this trend may be due to clusters that have just merged with a cool clump and are still on their way to virialization, as illustrated by the following example. Figure~\ref{hist_minor} is the time history constructed from our merger tree analysis for a cluster undergoing a minor merger. Here we plot the evolution of mass, temperature, and substructure
measures versus lookback time ($t=0$ for today). We can see that at $t = 1.3$\ Gyr,
a clump of cold gas merged into the primary cluster. The cold accreted gas caused a jump
in mass and substructure measures, but it reduced the average temperature of the cluster. 
This behavior thus tends to make merging clusters lie below the $M_{500}$--$T_{X,500}$ relation when they have just merged and then gradually move up when they become virialized.
 
\begin{figure*}[htbp]
\begin{center}
\epsscale{1.0}\plottwo{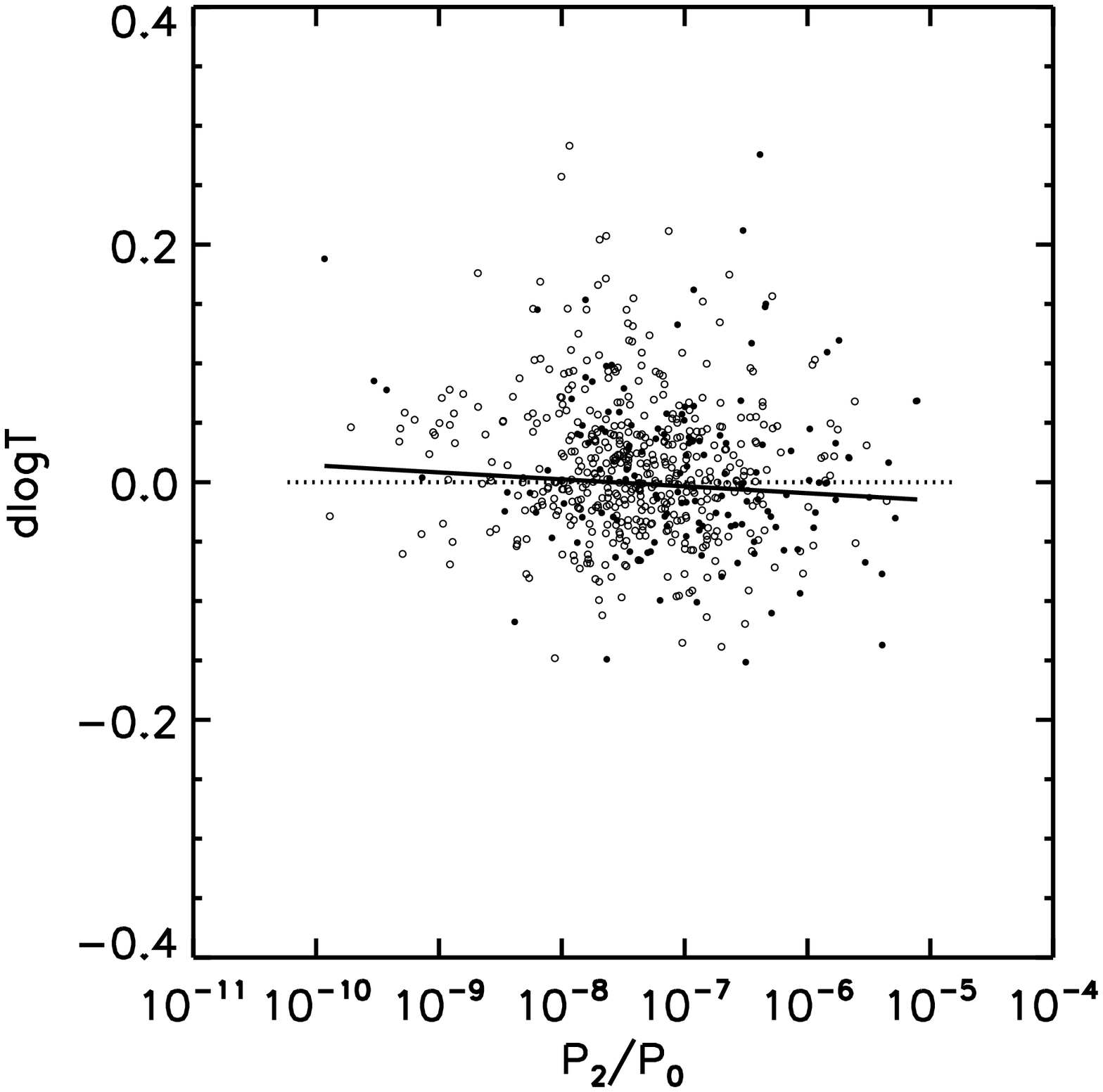}{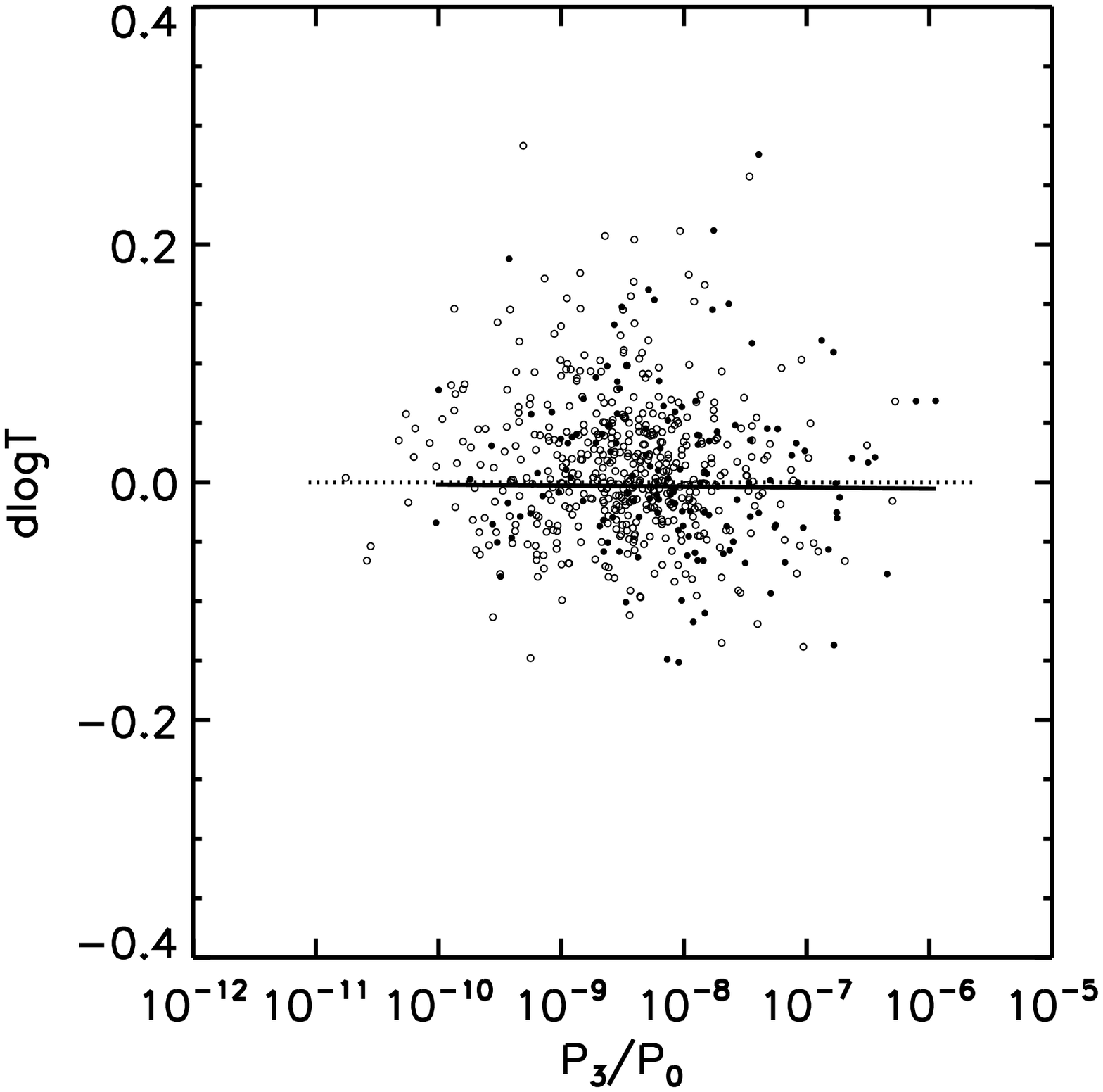}
\caption{$M_{500}$--$T_{X,500}$ scatter versus substructure measures $P_2/P_0$ and $P_3/P_0$. Relaxed/merging clusters are plotted in open/filled circles. Weak negative correlations are found for merging clusters with probabilities of no correlation being 0.08 for $P_2/P_0$ and 0.34 for $P_3/P_0$.}
\label{dt_sub}
\end{center}
\end{figure*}

This trend is also seen when we correlate scatter with some of the substructure indicators, although the probabilities are not high. Figure \ref{dt_sub} shows the correlation with the two substructure measures that give the highest probabilities, $P_2/P_0$ and $P_3/P_0$ measured with a fixed aperture radius 1\ Mpc. Here we use a fixed aperture size because this gives power ratios that are weaker functions of cluster mass than substructures computed using $R_{200}$ or $R_{500}$. This is to minimize covariance with halo concentrations through the mass dependence. This negative correlation between scatter and substructures again supports the idea that merging clusters tend to be cooler than clusters with similar masses.   

But do recent mergers cause a problem statistically? Again we want to compare the distributions of scatter for the merging and relaxed populations in the same way as in \S~\ref{Sec:scatter distribution}. But as discussed earlier, the trends are probably dominated by the effect of halo concentrations. Since we want to study the effect of recent mergers in isolation from other effects, we compute the statistics using the $M_{500}$--$T_{X,500}$ relation after correcting for halo concentrations (Figure \ref{mts_correctc}) instead of the raw relation. In this way we can see whether the dynamical state of clusters is the second dominant factor in the scatter.  

\begin{figure*}[htbp]
\begin{center}
\epsscale{1.0}\plottwo{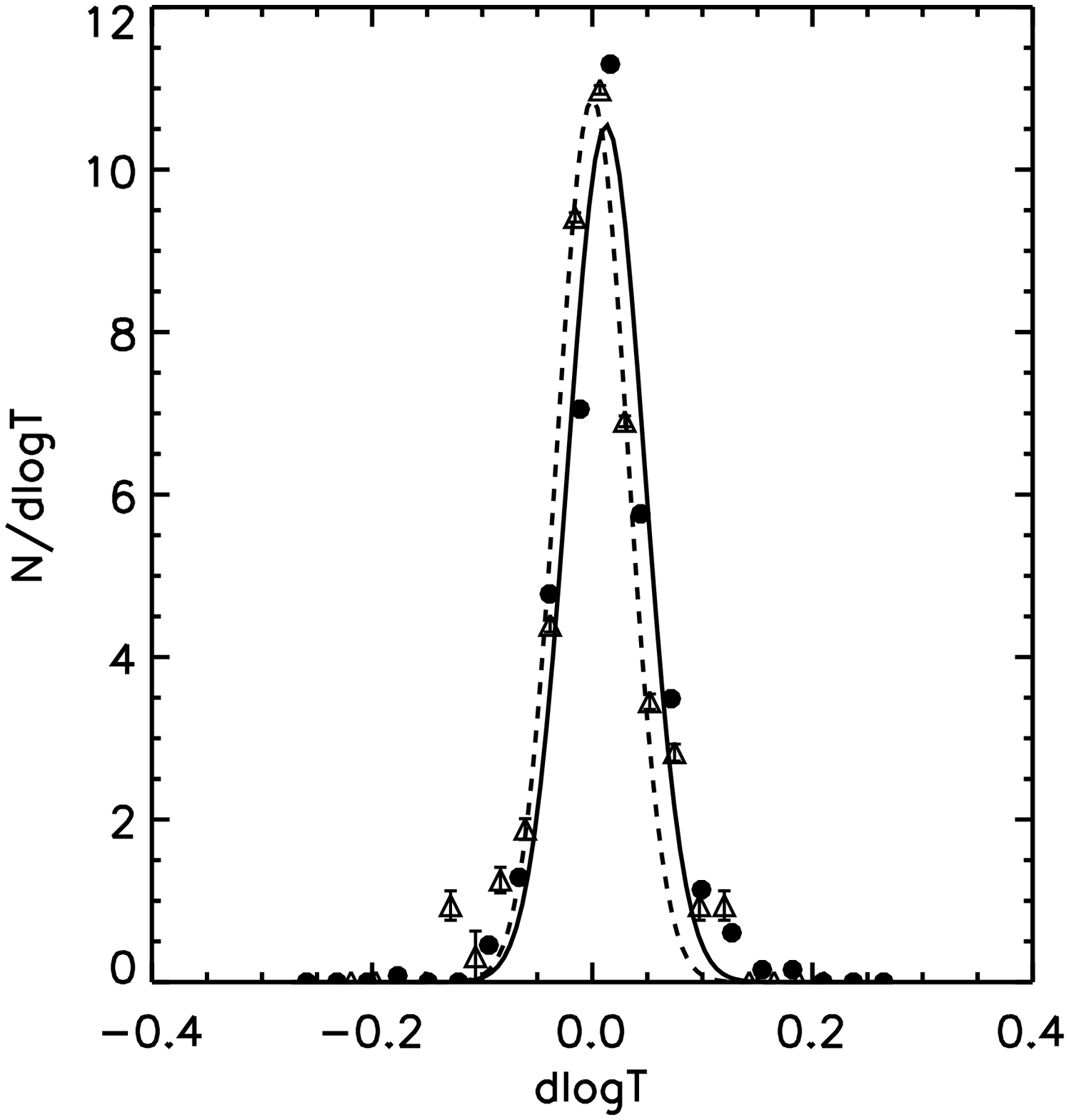}{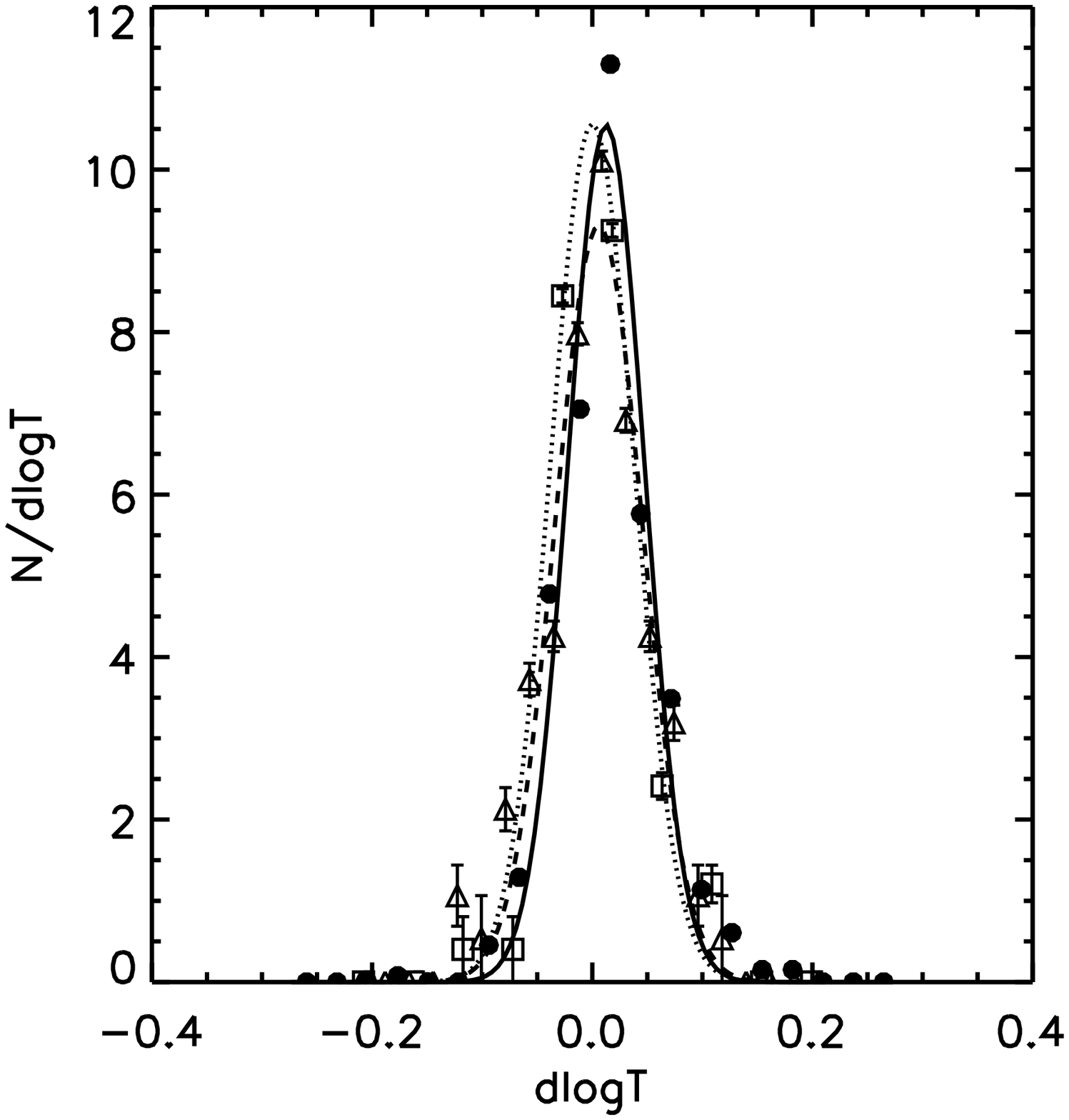}
\caption{Distributions of scatter at $z=0$ after correcting for the effect of halo concentrations. In the left panel, merging clusters are plotted using open triangles and dashed lines, and relaxed clusters are plotted using filled circles and solid lines. In the right panel, the merging clusters are separated into minor mergers (open squares and dotted lines) and major mergers (open traiangles and dashed lines).}
\label{mtsdev_dist_correctc}
\end{center}
\end{figure*}

The distributions of the concentration-corrected scatter for merging and relaxed clusters at $z=0$ are shown in Figure \ref{mtsdev_dist_correctc}. Again we use the R-S test and F-V test to detect whether there is any difference in the mean values and variances of these two populations. The results are summarized in Table \ref{sig_test_correctc}. We find that after removing the effect of halo concentration, the trend that merging clusters have a larger RMS scatter becomes insignificant. This supports our earlier statement that the behavior of the raw scatter is determined more by the distribution of halo concentrations than by the dynamical state of clusters. As for the mean values, there is a significant relative bias for merging clusters to have smaller means than relaxed clusters at $z=0$. This bias can be due to the incomplete virialization of merging clusters we just described. However, it is also possible that we have over-corrected for the halo concentrations for merging clusters in comparison with relaxed clusters, because merging clusters are generally less concentrated (or have larger $R_{200}/R_{500}$, see Figure \ref{c_dist}). 

In summary, by correlating the scatter with the dynamical state of clusters using both merger tree and substructure analysis, we find a very weak trend that merging clusters tend to be cooler than relaxed clusters with similar masses. But this trend has a minor effect on the statistical properties of the scatter, i.e., the scatter for the merging clusters is neither biased nor wider spread compared to the relaxed ones at $z=0$ and $z=1$ (see Table \ref{sig_test_correctc}). This implies that the dynamical state is not the second most important factor that contributes to the scatter, but that still other sources need to be found. However, the fact that the distributions of scatter for the merging and relaxed clusters are indistinguishable even out to higher redshift is good news for using cluster scaling relations in cosmology. 

\begin{table*}[thdp]
\caption{Significance tests on the distribution of scatter after removing the effect of halo concentrations for different populations at $z=0$ and $z=1$. 
The R-S and F-V test results for each subgroup are relative to the relaxed clusters.}
\begin{center}
\begin{tabular}{ccccccc}
\hline
\hline
Subgroup & $z$ & $N$ & Mean &  R-S Test & $\sigma_{rms} (\%)$ & F-V test \\
\hline
Relaxed & 0 & 478 & $1.44\times10^{-2}$ & - & 4.42 & -  \\ 
Merging & 0 & 141 & $4.27\times10^{-3}$ & 0.012 & 4.67 & 0.403  \\ 
Minor & 0 & 55 & $6.39\times10^{-3}$ & 0.058 & 4.63 & 0.612 \\ 
Major & 0 & 86 & $2.92\times10^{-3}$ & 0.034 & 4.72 & 0.406 \\ 
\hline
Relaxed & 1 & 102 & $1.03\times10^{-2}$ & - & 4.25 & - \\ 
Merging & 1 & 121 & $6.08\times10^{-3}$ & 0.329 & 4.88 & 0.151 \\
Minor & 1 & 46 & $1.30\times10^{-2}$ & 0.275 & 4.96 & 0.202 \\
Major & 1 & 75 & $1.86\times10^{-3}$ & 0.145 & 4.81 & 0.241 \\
\hline
\hline
\end{tabular}
\end{center}
\label{sig_test_correctc}
\end{table*}


\section{Discussion}
\label{Sec:discussion}

\subsection{Effect of dynamical state}
\label{Sec:dyn state}

In the previous section we have shown that the dynamical state of clusters has very little influence on the overall scatter in the $M_{500}$--$T_{X,500}$ relation. When we look at the merging population, there is even a tendency for merging clusters to be cooler than relaxed ones of similar masses. Although this can be explained by incomplete virialization of clusters merging with a cooler clump, it is still somewhat contrary to our intuition that merger shocks can heat the intracluster medium and raise the temperature of a merging cluster. We discuss the possible reasons for this result in the following.

\begin{figure*}[pthb]
\begin{center}
\includegraphics[height=7.25in]{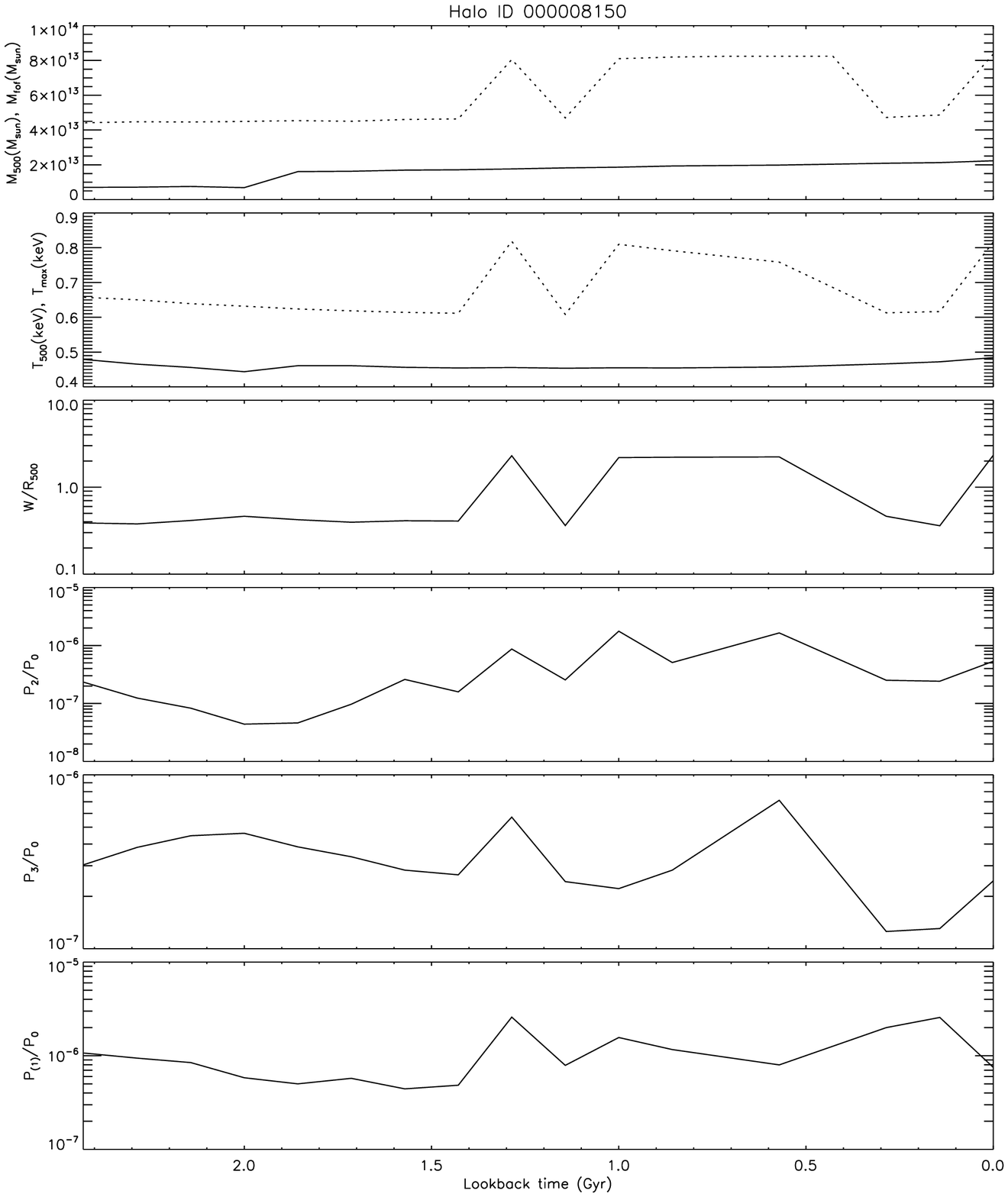}
\caption{Example time history of a cluster that undergoes a major merger. The solid (dashed) line 
is $M_{500}$ ($M_{fof}$) in the first panel, $T_{500}$ ($T_{max}$) in the second panel. The
other panels show the evolution of the substructure measures.}
\label{hist_major}
\end{center}
\end{figure*}

One reason is that sometimes the merger shock is not captured by the projected $R_{500}$ aperture. At the beginning of mergers, the shocks often occur in the outskirts of clusters. So in order for the shock to be captured inside $R_{500}$, either the shock has to propagate into the $R_{500}$ region of the main cluster, or the two clusters have to merge roughly along the line of sight in order to affect $T_{X,500}$. The cluster history shown in Figure \ref{hist_major} is one example of such a case. At $t = 1.3$\ Gyr, the substructure measures increase, indicating the start of the merger event. The maximum temperature in the cluster is increased by the merger shock, but $T_{X,500}$ is unaffected because the shock-heated gas lies beyond the projected aperture radius $R_{500}$. 
Even if the merger shock is within $R_{500}$, the spectroscopic temperature is not as sensitive to shocks as the emission-weighted temperature because when the shock and other cooler gas in the cluster are projected along the line of sight, the spectral fit tends to put more weight on the cooler gas \citep{Mazzotta:2004}.

Secondly, the duration of the temperature boost is typically only $\sim0.5$ Gyr \citep{2001ApJ...561..621R, 2007MNRAS.380..437P}, and hence only a fraction of the unrelaxed clusters are observed during the transient excursion period. Moreover, even for those clusters that have just undergone major mergers, the increase in mass and temperature are often comparable. For example, a 3:1 merger would have a mass jump of $M_f/M_i \sim 1.33$ and a temperature jump of $T_f/T_i \sim 2$ depending on the impact parameter of collision. Thus clusters tend to evolve roughly parallel to the scaling relation, as also suggested by previous works \citep[e.g.][]{2007MNRAS.380..437P}. 

Finally, merging clusters are the minority population compared to relaxed clusters. The fraction of clusters that had a merger within the past 3\ Gyr is $\sim 23\%$ at $z=0$ and $\sim 54\%$ at $z=1$, while major mergers are rarer. All these effects combined are responsible for diluting the influence of mergers on the scatter and making their distributions indistinguishable from relaxed clusters.

\subsection{Effectiveness of substructure indicators}

\begin{figure*}[htbp]
\begin{center}
\epsscale{1.0}\plottwo{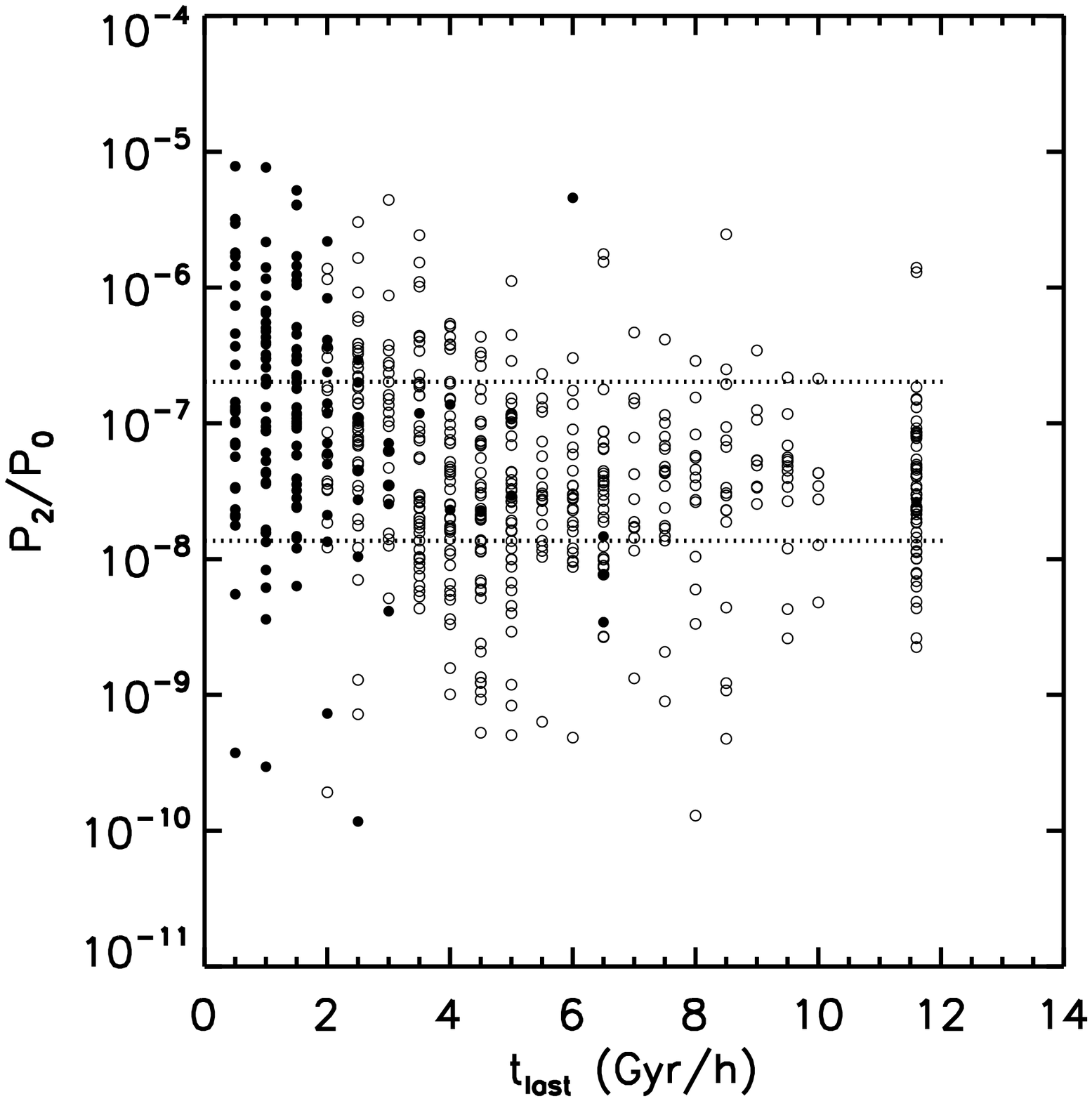}{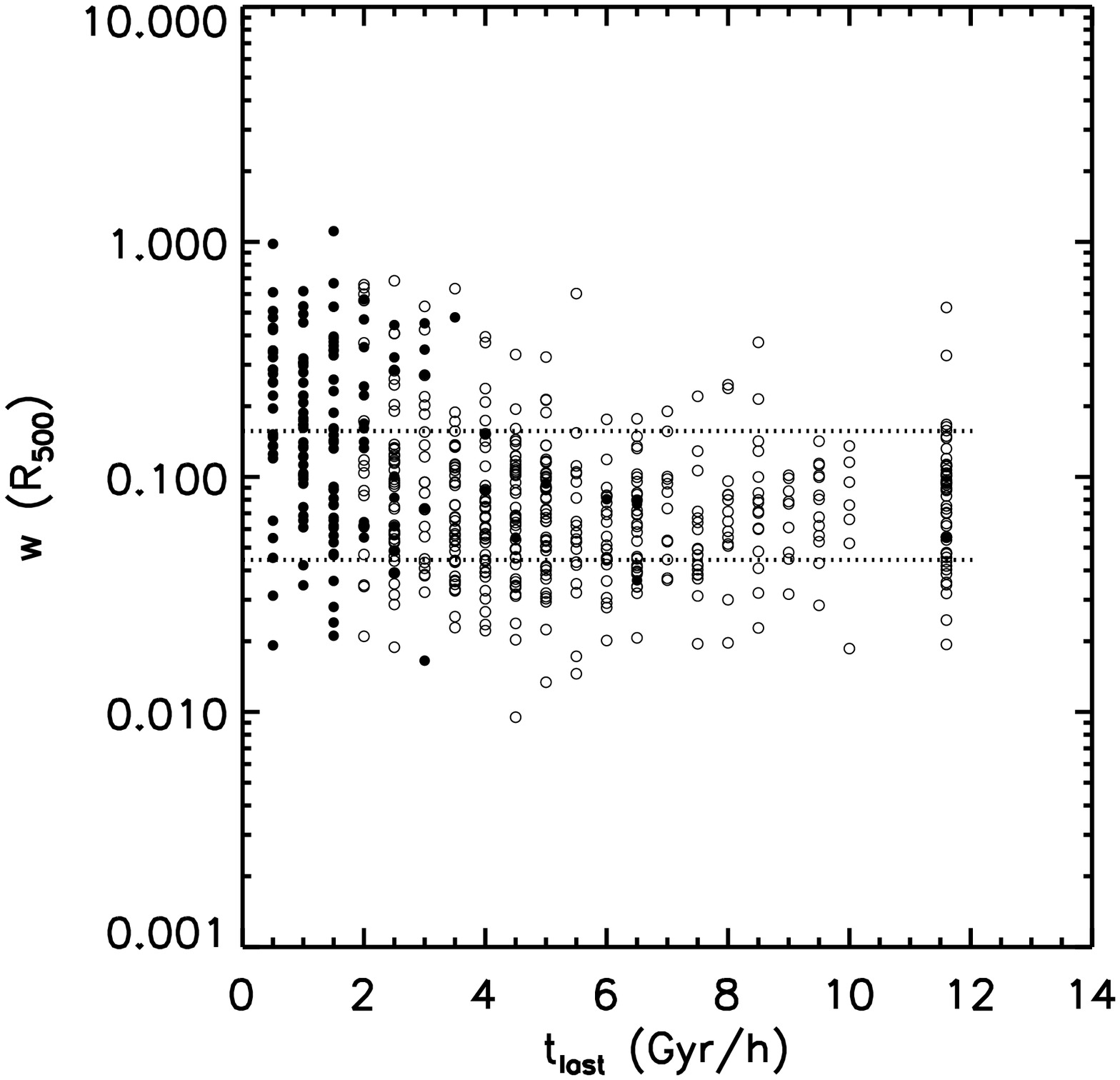}
\caption{Substructure measures $P_2/P_0$ and $w$ in x projection versus the time since the last $\leq 5:1$ merger. The filled and open circles are merging and relaxed clusters identified by the merger tree diagnostic, respectively. The two horizontal dashed lines mark the upper and lower 20\% of all clusters that have the highest and lowest substructure values. The right most column in the figure are clusters that did not have any $\leq 5:1$ merger in their entire histories.}
\label{sub_tlast}
\end{center}
\end{figure*}

Substructure measures, such as centroid offsets and power ratios, have often been used in earlier studies to identify unrelaxed clusters \citep[e.g.][]{2006ApJ...639...64O, 2007MNRAS.377..317K, 2008ApJ...681..167J}. Observationally, they are quantities that link most effectively to the dynamical state of clusters \citep{1993ApJ...413..492M, 1996ApJ...458...27B}. In simulations, they are easy to derive and to compare with observations. However, they still have some limitations, such as the projection effect.

In our study we have extended the analysis of the dynamical state of clusters by constructing cluster merging histories, because these provide more information about the true dynamics of clusters during mergers than morphology-based measures, which are subjected to observational limitations.
This is the first time that these two approaches have been compared to see how effectively the substructure measures can recover the true merging population. 

Figure \ref{sub_tlast} shows the negative correlations between two of the substructure measures as seen in the x projection of the simulation box and the time since the last $\leq 5:1$ merger for $z=0$. The filled and open circles are merging and relaxed clusters identified by the merger tree diagnostic, respectively. The two horizontal dashed lines mark the upper and lower 20\% of all clusters that have the highest and lowest substructure values. We can see that, indeed, recently merged clusters tend to have more substructure than relaxed clusters. However, the merging and relaxed populations overlap over a wide range of substructure values because of the large variation in substructures even for clusters at the same dynamical state. Therefore, there is not a clean cut to separate these two populations using the substructure measures. 

In order to facilitate future studies in this area, we compute the completeness and contamination of merging and relaxed clusters defined by the substructure measures, assuming those identified using the cluster merger histories represent the true populations. We refer by `completeness' to the fraction of merging/relaxed clusters found using substructure among the true merging/relaxed population, while `contamination' is the fraction of clusters that are detected as merging/relaxed but are actually relaxed/merging. In Table \ref{sub_effectiveness} we summarize the results for different substructure measures. In the definitions using substructures, clusters that lie above/below the 20\% thresholds are defined as merging/relaxed. In order to see how much the results are affected by the projection effect, 
we list the values obtained using information from both one and all projections. In the latter case, we select clusters by comparing the maximum value among three projections to the selection thresholds. In other words, clusters are identified as merging if the most disturbed value among three projections is above the 20\% threshold, and relaxed clusters must have their most disturbed value below the lower 20\% threshold.  

By comparing the numbers in Table \ref{sub_effectiveness}, we find that in general $P_2/P_0$ and $P_3/P_0$ give similar results, and $P_{(1)}/P_0$ is not as useful as the other two power ratios. The centroid offset, having the greatest completeness and the least contamination, is the most successful one among all measures. When different aperture sizes are compared for the power ratios, the results using aperture sizes of 1\ Mpc and $R_{200}$ are similar, while using $R_{500}$ is generally a little worse than the others both in completeness and contamination. When all three projections are considered, merging and relaxed clusters are better distinguished for all measures, with the values of completeness and contamination changed by $\sim 10-20\%$. The centroid offset, $w$, improves the most when all projections are used. 

According to the above analysis, the centroid offset does a better job in distinguishing merging and relaxed clusters than the power ratios. This is probably because each individual power ratio is only sensitive to a certain type of substructure. They are more powerful when combined to distinguish between different morphological types \citep{1995ApJ...452..522B, 1996ApJ...458...27B}. The centroid offset, on the other hand, is a more general feature of all disturbed clusters. However, all the substructure measures have limitations. Their effectiveness is influenced by the viewing projection. More importantly, their values have a large variation even for clusters at the same dynamical state, as shown in Figure \ref{sub_tlast}. Thus only $\sim 40\%$ of the true merging clusters are detected, and $\sim 25\%$ of the relaxed clusters. Among the detected clusters, $\sim 50\%$ of the ``merging'' clusters are actually relaxed, and $\sim 15\%$ of the ``relaxed'' clusters are actually merging. Therefore, although substructure measures are useful in distinguishing the dynamical state of clusters, caution is still required to interpret the results correctly.

\begin{table*}[thdp]
\caption{Completeness and contamination of merging and relaxed clusters identified using power ratios and centroid offsets. Clusters lying above/below the 20\% thresholds are defined as merging/relaxed. For the upper half of the table, the selection is based on the values of substructure measures in one projection, while for  the bottom half, clusters are found by comparing the maximum value among three projections to the thresholds. See the text for details on the definition of completeness and contamination.}
\begin{center}
\begin{tabular*}{\textwidth} {@{\extracolsep{\fill}} l|ccc|ccc|ccc|c}
\hline
\hline
&  \multicolumn{3}{c|}{$P_2/P_0$} & \multicolumn{3}{c|}{$P_3/P_0$} & \multicolumn{3}{c|}{$P_{(1)}/P_0$}  & $w$ \\ [+5pt]
& {\scriptsize(1Mpc)} & {\scriptsize($R_{200}$)} & {\scriptsize($R_{500}$)} & 
    {\scriptsize(1Mpc)} & {\scriptsize($R_{200}$)} & {\scriptsize($R_{500}$)} & 
    {\scriptsize(1Mpc)} & {\scriptsize($R_{200}$)} & {\scriptsize($R_{500}$)} & 
    {\scriptsize($R_{500}$)} \\ [+2pt]
\hline
Comp.\ of merging & 39.6 & 38.2 & 31.3 & 36.8 & 34.7 & 28.5 & 22.2 & 21.5 & 23.6 & 47.9 \\ [+2pt]
Cont.\ of merging & 54.0 & 55.7 & 63.7 & 57.3 & 59.7 & 66.9 & 74.2 & 75.0 & 72.6 & 44.3 \\ [+2pt]
Comp.\ of relaxed & 22.8 & 23.2 & 21.8 & 22.2 & 23.4 & 21.6 & 19.3 & 19.3 & 18.0 & 23.9 \\ [+2pt]
Cont.\ of relaxed & 12.8 & 11.2 & 16.8 & 15.2 & 10.4 & 17.6 & 26.4 & 26.4 & 31.2 & 8.94 \\
\hline
& \multicolumn{3}{c|}{$(P_2/P_0)_{max}$} & \multicolumn{3}{c|}{$(P_3/P_0)_{max}$} & \multicolumn{3}{c|}{$(P_{(1)}/P_0)_{max}$} & $w_{max}$ \\ [+3pt]
\hline
Comp.\ of merging & 40.3 & 44.4 & 33.3 & 42.4 & 37.5 & 29.9 & 25.7 & 25.0 & 22.9 & 54.9 \\ [+2pt]
Cont.\ of merging & 53.2 & 48.4 & 61.3 & 50.8 & 56.5 & 65.3 & 70.2 & 71.0 & 73.4 & 36.1 \\ [+2pt]
Comp.\ of relaxed & 21.6 & 23.2 & 22.0 & 22.8 & 22.8 & 22.2 & 20.1 & 19.5 & 21.6 & 25.0 \\ [+2pt]
Cont.\ of relaxed & 17.6 & 11.2 & 16.0 & 12.8 & 12.8 & 15.2 & 23.2 & 25.6 & 17.6 & 4.9 \\
\hline
\hline
\end{tabular*}
\end{center}
\label{sub_effectiveness}
\end{table*}

\subsection{Comparison with previous work}

Several studies have explored the effect of clusters with different dynamical states on the $M-T_X$ scaling relation. Most previous works used substructure measures such as the power ratios and the centroid offset to quantify the dynamical state. The main difference in our study is that we directly analyze cluster merging histories to identify the recently merged clusters, which provides another line of evidence for our results in addition to those derived from the substructure measures.

\cite{2006ApJ...639...64O} investigated the effect of mergers and core structure on the X-ray scaling relations for both observed and simulated clusters. For the observed sample, they found that cool core clusters and clusters with less substructure exhibit a larger amount of scatter. Their simulated clusters, on the other hand, have a tendency to have a larger amount of scatter for clusters with more substructure, though they argued that the evidence is weak. Since their simulations also do not include radiative cooling, we can compare directly with their results without worrying about other baryonic effects. We also find the same trend that merging clusters, which we have shown to have more substructure, have a larger amount of scatter. However, we further explore the origin of the intrinsic scatter and find a strong correlation with the halo concentration of clusters. We also show that the trend seen above is due to the fact that merging clusters have a larger variation in their concentrations.  

\cite{2008ApJ...681..167J} studied the correlation between cluster substructures and cluster observables using hydrodynamical simulations with non-gravitational heating and cooling. Despite the difference of input baryonic physics in the simulations and the definition of $T_X$, they found no dependence on cluster substructures in the $M-T_{ew}$ relation when the true mass is considered. Although we find negative correlations between the scatter and some of the substructure indicators, the probabilities are not high. The lack of correlation supports their result. However, they reported that there is a significant trend for the relaxed clusters to have lower temperatures for their masses in the $M-T_{ew}$ relation measured within $R_{500}$, whereas we do not find any significant bias, and an opposite trend is found by \cite{2006ApJ...650..128K}, \cite{2007ApJ...668....1N} and \cite{2007MNRAS.377..317K}. Taking the average relations between $T_{ew}$ and $T_X$ for merging and relaxed clusters from \cite{2007ApJ...655...98N}, \cite{2008ApJ...681..167J} argued that the discrepancy cannot be explained by using different temperature definitions. However, we find that the bias between $T_{ew}$ and $T_X$ is larger for merging clusters than relaxed ones (see \S~\ref{Sec:xray}), in the direction that can alleviate this discrepancy. Therefore, the difference between the conclusions reached by \cite{2008ApJ...681..167J} and the others regarding the offset of merging and relaxed clusters may be due to the use of $T_{ew}$ instead of the spectroscopic temperature $T_X$.

On the other hand, \cite{2008ApJ...685..118V} have recently found significant negative correlations between the substructure measures and the scatter in the mass-temperature relation for their simulated clusters, both for the emission-weighted temperature $T_{ew}$ and the spectroscopic-like temperature $T_{sl}$. They also found relative offsets between merging and relaxed clusters in the mean scaling relation, that is, merging clusters tend to be cooler than relaxed clusters of similar masses. As discussed in their paper, the different conclusion than \cite{2008ApJ...681..167J} may come from different implementations of feedback mechanisms. Although the trends can be explained by incomplete relaxation of merging clusters, which we also observed, we do not find a significant separation in the normalization of the $M-T_X$ relation between merging and relaxed populations, especially when the intrinsic scatter without cooling is largely contributed by the effect of halo concentration. Therefore, the main reason resulting in the differences is probably due to radiative cooling, as is suggested by the fact that it is included in all the studies which observed the relative offset. Since current simulations with radiative cooling tend to produce relaxed clusters with steeper temperature profiles than real clusters, the average temperature of relaxed clusters can possibly be biased high. This can explain why they found relaxed clusters hotter than expected while our results do not show any significant offset between relaxed and merging clusters in the $M-T_X$ relation. 


\section{Conclusions}
\label{Sec:conclusion}

Galaxy clusters are invaluable cosmological probes. Accurate measurement of cluster masses is crucial and often relies on the mass-observable relations. However, to constrain the cosmological parameters to the few percent level, the systematics and scatter in these relations must be thoroughly understood. In this work we investigate the sources of intrinsic scatter in the $M-T_X$ relation using a hydrodynamics plus N-body simulation of galaxy clusters in a cosmological volume. In order to compare directly to observations without worrying about different definitions or systematics, we produced mock Chandra X-ray images using MARX and extracted the spectroscopic temperature $T_X$ as observers do. Also, all the quantities in our analysis and discussions are measured within $R_{500}$, which is usually used for X-ray data.
Radiative cooling and heating mechanisms are not included, since we would like to disentangle the scatter driven by the gravitational effects from other baryonic physics.
We chose to focus on the $M-T_X$ relation for several reasons. First, it is less sensitive to resolution and to cooling and heating mechanisms than other scaling relations, such as the $L_X-T_X$ relation. Therefore, our results are still representative of reality even though the input physics is not complete. Second, 
the relative insensitivity to additional baryonic physics provides a window into better understanding of the physical origin of the scatter, despite our incomplete knowledge regarding the cooling and feedback mechanisms. Moreover, the intrinsic scatter in the $M-T_X$ relation is among the smallest of all the observed scaling relations. If one can further reduce it based on the knowledge of its physical origin, the $M-T_X$ relation will be extremely useful for cluster cosmology. 

Our aim is to find out what determines the positions of clusters in the $M_{500}$--$T_{X,500}$ relation, in particular whether the intrinsic scatter is driven by recent merging activity or the overall assembly histories of clusters. We split our simulated cluster samples into merging and relaxed subgroups based on our merger tree analysis, and then we compare the distributions of the intrinsic scatter for individual subgroups. We also correlate the scatter with quantities that are related to the recent merging activity or cumulative cluster assembly histories, including the time since last merger, substructure measures, and the halo concentration. Here we summarize our findings.

We find a strong correlation between the scatter in the $M_{500}$--$T_{X,500}$ relation and the halo concentration. More concentrated clusters tend to lie below (cooler than) the mean relation, while puffier clusters tend to be hotter than expected for their masses. This is confirmed by the negative correlation between scatter and the formation lookback time of clusters, since it is well known that more concentrated clusters tend to form at earlier times. We showed that using this correlation, the scatter can be effectively reduced from $6.10\%$ to $4.49\%$. 
 
There is no bias in the $M_{500}$--$T_{X,500}$ relation between merging and relaxed clusters, but the amount of scatter for merging clusters is larger than that for the relaxed ones. This trend can be explained by the fact that merging clusters have larger variations in their concentrations than relaxed clusters. 

When we correlate the scatter with the dynamical state of clusters, either the time since last merger or the substructure measures, there is a weak trend for recently merged clusters to be cooler, probably due to incomplete virialization of clusters that have just merged with a cool clump. However, statistically the influence of departure from hydrostatic equilibrium of merging clusters is negligible. Possible reasons are discussed in detail in \S~\ref{Sec:dyn state}.  

There are significant deviations from lognormality of the distributions of scatter for our simulated clusters at both $z=0$ and $z=1$. This effect should be taken into account in future self-calibration studies to correctly interpret the obtained constraints on the cosmological parameters. 
Future simulation studies of larger volumes are needed in order to accurately characterize the distribution of scatter, including the tails of the distribution. 

In conclusion, we find that when radiative cooling and feedback mechanisms are neglected, the intrinsic scatter in the $M_{500}$--$T_{X,500}$ relation is driven more by the variation in halo concentrations, or the overall assembly histories of clusters, than the recent merging events. Using an analytic approach, \cite{2006MNRAS.366..624B} investigated whether the amount of scatter in the observed $M-T_X$ and $M-L_X$ relations can be explained by the variations in halo concentrations or different entropy floors. Although they focused more on the scatter in the $M-L_X$ relation and showed that it requires a wide range of entropy floors, a significant portion of the scatter in the $M-T_X$ relation is determined by the range of halo concentrations predicted in their model. This is confirmed by our results since we have explicitly shown the strong correlation between the $M-T_X$ scatter with the halo concentration using a numerical simulation. 

The lack of dependence of the scatter on the dynamical state of clusters is also seen by \cite{2006ApJ...639...64O} for observed clusters. They suggested that the scatter in the mass-observable relations is not dominated by recent mergers, but by the cooling-related core properties or probably the overall assembly histories of clusters. The latter relationship is indeed found in our simulation. As for the effects of radiative cooling, it has been shown to be a main source of intrinsic scatter in the $L_X-T_X$ relation \citep{1998MNRAS.297L..57A}. Therefore, our next step is to include cooling and feedback mechanisms in the simulation to examine their individual influence on the scatter in the mass-observable relations.

Exploring the intrinsic scatter in the cluster scaling relations not only provides physical insights into the formation of galaxy clusters, but also has important implications for using clusters in cosmology. For example, the strong correlation with halo concentrations can be used for observed clusters to reduce the scatter in the scaling relations. Also, the weak influence of merging clusters is good news for cluster cosmology, because it implies that when deriving the scaling relations from the observed clusters, it is unnecessary to worry much about whether to include the unrelaxed systems or not. This is good because it is much more difficult to separate out the unrelaxed systems at higher redshifts. We expect that detailed studies of the intrinsic scatter in the scaling relations, not only in X-rays but also at other wavelengths, will continue to yield invaluable information both for cluster physics and cluster cosmology.  


\acknowledgments

HYY wishes to thank J.\ Cohn for hospitality during a visit to Berkeley and J.\ Zuhone for useful conversations. PMS acknowledges support from the DOE Computational Science Graduate Fellowship
(DEFG02-97ER25308).
We acknowledge support under a Presidential Early Career Award
from the U.S. Department of Energy, Lawrence Livermore National Laboratory
(contract B532720) and from NASA (grant NNX06AG57G).
The work described here was carried out using the resources of the
National Center for Supercomputing Applications (allocation MCA05S029)
and the National Center for Computational Sciences at Oak Ridge
National Laboratory (allocation AST010).
FLASH was developed largely by the
DOE-supported ASC/Alliances Center for Astrophysical Thermonuclear Flashes at
the University of Chicago.


\bibliography{scatter}

\end{document}